\documentclass[letterpaper]{article}
\usepackage[T1]{fontenc}
\usepackage{aaai}
\usepackage{times}
\usepackage{helvet}
\usepackage{courier}

\usepackage{subfigure}
\usepackage{graphicx,url}
\usepackage{fancyhdr}
\usepackage{rotating}
\usepackage{subfigure}
\usepackage{amssymb}
\usepackage{amsmath}
\usepackage{multirow}
\usepackage{multicol}
\usepackage{indentfirst}
\usepackage{float}
\usepackage{epsfig}
\usepackage{algorithm,algorithmic}
\usepackage{color}

\newcommand{\remove}[1]{}

\begin{document}

\title{Friendship Paradox Redux: Your Friends Are More Interesting Than You}
\author{
Nathan O. Hodas \\
USC Information Sciences Institute\\
4676 Admiralty Way \\
Marina del Rey, CA 90292 \\
\texttt{nhodas@isi.edu} \\
\And
Farshad Kooti \\
USC Information Sciences Institute\\
4676 Admiralty Way \\
Marina del Rey, CA 90292 \\
\texttt{kooti@usc.edu} \\
\And
Kristina Lerman \\
USC Information Sciences Institute\\
4676 Admiralty Way \\
Marina del Rey, CA 90292 \\
\texttt{lerman@isi.edu}
}
\maketitle

\begin{abstract}
Feld's friendship paradox states that ``your friends have more friends than you, on average.'' This paradox arises because extremely popular people, despite being rare, are overrepresented when averaging over friends.  Using a sample of the Twitter firehose, we confirm that the friendship paradox holds for >98\% of Twitter users.  Because of the directed nature of the follower graph on Twitter, we are further able to confirm more detailed forms of the friendship paradox:  everyone you follow or who follows you has more friends and followers than you.  This is likely caused by a correlation we demonstrate between Twitter activity, number of friends, and number of followers.  In addition, we discover two new paradoxes: the \emph{virality paradox} that states ``your friends receive more viral content than you, on average,'' and the \emph{activity paradox}, which states ``your friends are more active than you, on average.''  The latter paradox is important in regulating online communication.
It may result in users having difficulty maintaining optimal incoming information rates, because following additional users causes the volume of incoming tweets to increase super-linearly.
While users may compensate for increased information flow by increasing their own activity,  users become information overloaded when they receive more information than they are able or willing to process. We compare the average size of cascades that are sent and received by overloaded and underloaded users. And we show that overloaded users post and receive larger cascades and they are poor detector of small cascades.
\end{abstract}

\section{Introduction}

\noindent The so-called ``Friendship Paradox" or Feld's Paradox, states that, on average, your friends have more friends than you do. This is due to the overrepresentation of extremely popular individuals in the  average of friends~\cite{Feld91}.  The paradox has been empirically demonstrated both online, such as Facebook~\cite{Ugander11}, and offline~\cite{Feld91,zuckerman2001makes} social networks.
Because people use their local network to assess themselves and as sources of information about the greater world~\cite{zuckerman2001makes,sgourev2006lake,wolfson2000students,yoganarasimhan2012impact,kanai2012brain}, the  friendship paradox leads to systematic biases in our perceptions.
For example,  a majority of people believe they possess above average driving skill~\cite{McKenna199145,groeger1989assessing}.
Furthermore, many personal characteristics correlate with high network degree, such as
the incidence of drug and alcohol use~\cite{tucker2011substance,tucker2012temporal}, wealth~\cite{morselli2004criminal,amuedo2007social,van2003network}, and extraversion~\cite{pollet2011extraverts,quercia2012personality}, which may further effect our perceptions.
Interestingly, your friends' superior social connectivity puts them at a greater risk, in aggregate, of an infection by a  biological pathogen.
This fact has been used as a principle for establishing epidemiological early-warning networks, because your friends will be more heavily exposed to pathogens in aggregate~\cite{Christakis10}.
Managing one's social network requires cognitive effort, which has been linked directly to physiological attributes within the brain~\cite{dunbar1993coevolution,powell2012orbital,bickart2012intrinsic,kanai2012online}.  However, the effect of the friendship paradox on our cognitive limitations is not well examined.



In online social networks, the friendship paradox has a surprising twist. If we wish to receive more information, we can usually choose to incorporate more individuals into our online social networks, e.g., by following them on Twitter.  However, as we grow our social network, we dramatically increase the volume of incoming information, since, as we show in this paper, not only are your friends better connected than you, they also tend to be more active, producing more information on average than you are willing to consume.
Thus, increase in information flow collides with our innate cognitive limitations and does not increase our ability to appreciate the totality of our relationships.
By increasing the incoming flow of information, we dilute our attention and reduce the visibility of any individual tweet~\cite{Hodas12socialcom}.  Receiving too much information may exceed our ability and desire to maintain existing social connections, even if they are unreciprocated~\cite{kwak2011fragile}.  Thus, users will naturally attempt to regulate the amount of incoming information by tuning the number of users they follow.


 In the present work, we consider the evidence for and the consequences of the friendship paradox on Twitter, which, as a directional network, presents an opportunity to study the paradox in  more detail.
 In the first part, we demonstrate the present evidence that the friendship paradox holds, as expected, on Twitter.
 We expand this analysis to other properties of the friendship network, presenting a full reciprocity friendship paradox:  your friends (followees) and followers have more friends and followers than you do.
 We then document new behavioral paradoxes.  The friend \emph{activity paradox} states that  your friends tend to be more active than you are. 
 Thus, the behavioral traits that lead one to be well connected will also have direct influence on information overload.
 Your friends also send and receive content that has higher virality than you do, what we call the \emph{virality paradox}.  These facts together suggest the glib expression ``your friends are more interesting than you are''.
 In the second part, we explore how the relative information load caused by the activity paradox alters user behavior, comparing underloaded users with overloaded users. 
 We show that, compared to underloaded users, overloaded users both post and receive more viral URLs and are less sensitive to smaller outbreaks of less popular URLs.

\section{A Variety of Paradoxes on Twitter} \label{sec:paradox}

\remove{
The friendship paradox was observed in offline world by Feld [ref] and also recently found
in Facebook[ref]. In both cases, the relationship between two friends is reciprocal.
But, in Twitter the reciprocity doesn't necessarily exist. In this section, we show
that the friendship paradox also exists in a directed friend graph like Twitter.
}

\noindent The friendship paradox, as formulated by Feld, is applicable to offline relationships, which are undirected, and it has also been observed in the undirected social network of Facebook~\cite{Ugander11}. We demonstrate empirically that the friend paradox also exists on Twitter. Unlike the friendship relations of the offline world and Facebook, the relations on Twitter are directed. When user $a$ follows the activity of user $b$, he or she can see the posts tweeted by $b$ but not vice versa. We refer to user $a$ as the \emph{follower} of $b$, and $b$ as a \emph{friend} or followee of $a$. Note that here friendship is a directed relationship.

\begin{figure}[tb]
\begin{center}
\includegraphics[width=0.8\columnwidth]{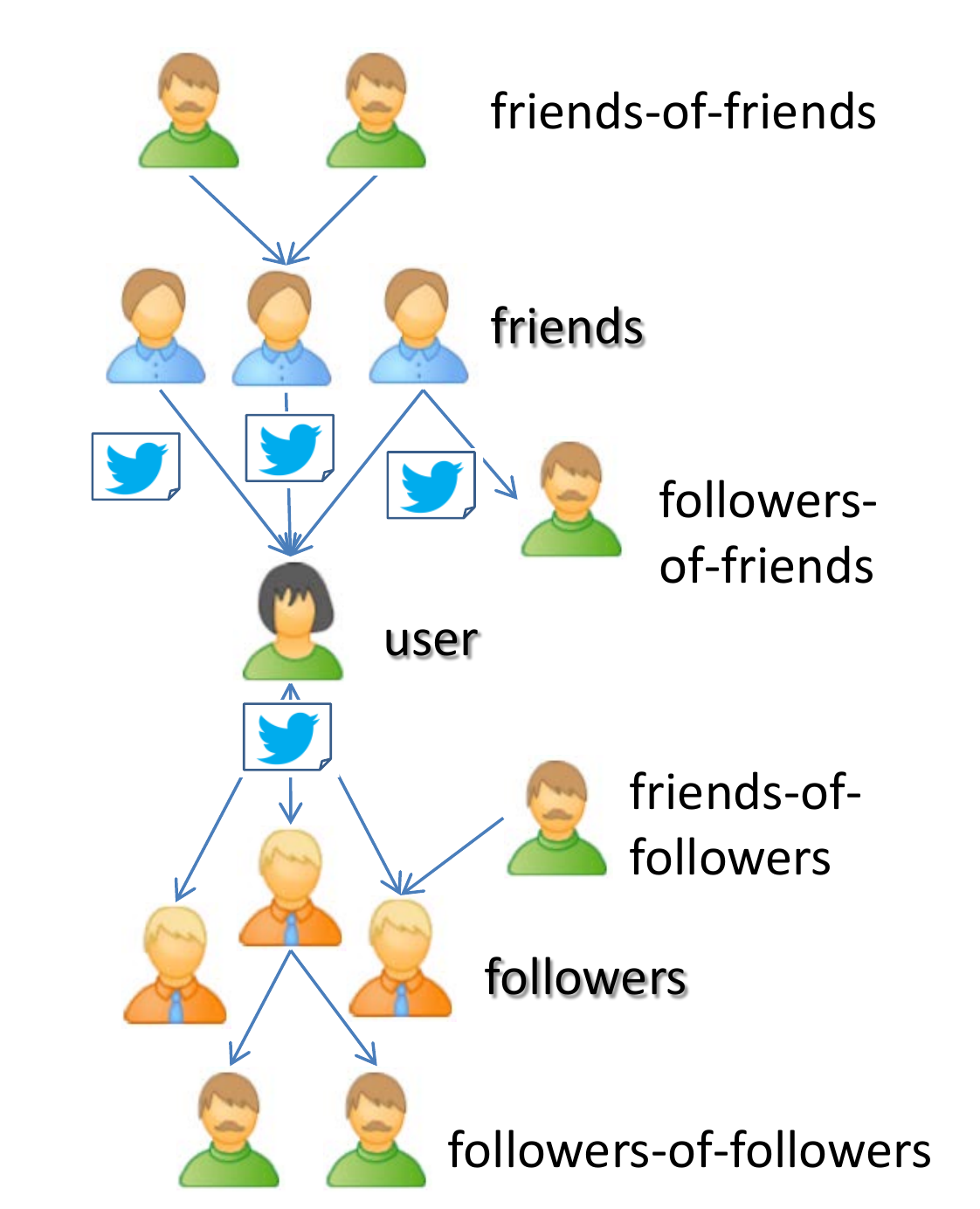}
\caption{\it An example of a directed network of a social media site with information flow links. Users receive information from their friends and broadcast information to their followers.}
\label{fig:directed}
\end{center}
\end{figure}

Figure~\ref{fig:directed} illustrates a directed social network of a social media site, such as Twitter. The \emph{user} receives information from \emph{friends} and, in turn, posts information to  her or his \emph{followers}. The friends may themselves receive broadcasts from their  friends, whom we call \emph{friends-of-friends} and post tweets to their own followers, whom we call \emph{followers-of-friends}.

\subsection{Data}
We use the Twitter dataset presented by~\cite{Yang2011wsdm}, which contains 476 million
tweets that are 20-30\% of all tweets posted from June to December 2009.
We also used the Twitter social network gathered by Kwak et al. (2010), which
includes links between all users who joined Twitter before August 2009.
Since we need both tweets and social links, we only consider  users who have posted
at least one tweet. The subgraph of such users includes
5.8M users and 193.9M links between them. This graph is used for showing the friendship
paradox on Twitter.

\subsection{Friendship Paradox}
The friendship paradox can be stated in four different ways on a directed graph:
\begin{itemize}

\item[$i$)] \emph{On average, your friends (followees) have more friends than you do.}
\item[$ii$)] \emph{On average, your followers have more friends than you do.}
\item[$iii$)] \emph{On average, your friends have more followers than you do. }
\item[$iv$)] \emph{On average, your followers have more followers than you do.}
\end{itemize}

\begin{figure*}[tbh]
\begin{center}
\begin{tabular}{cccc}
\includegraphics[width=0.5\columnwidth]{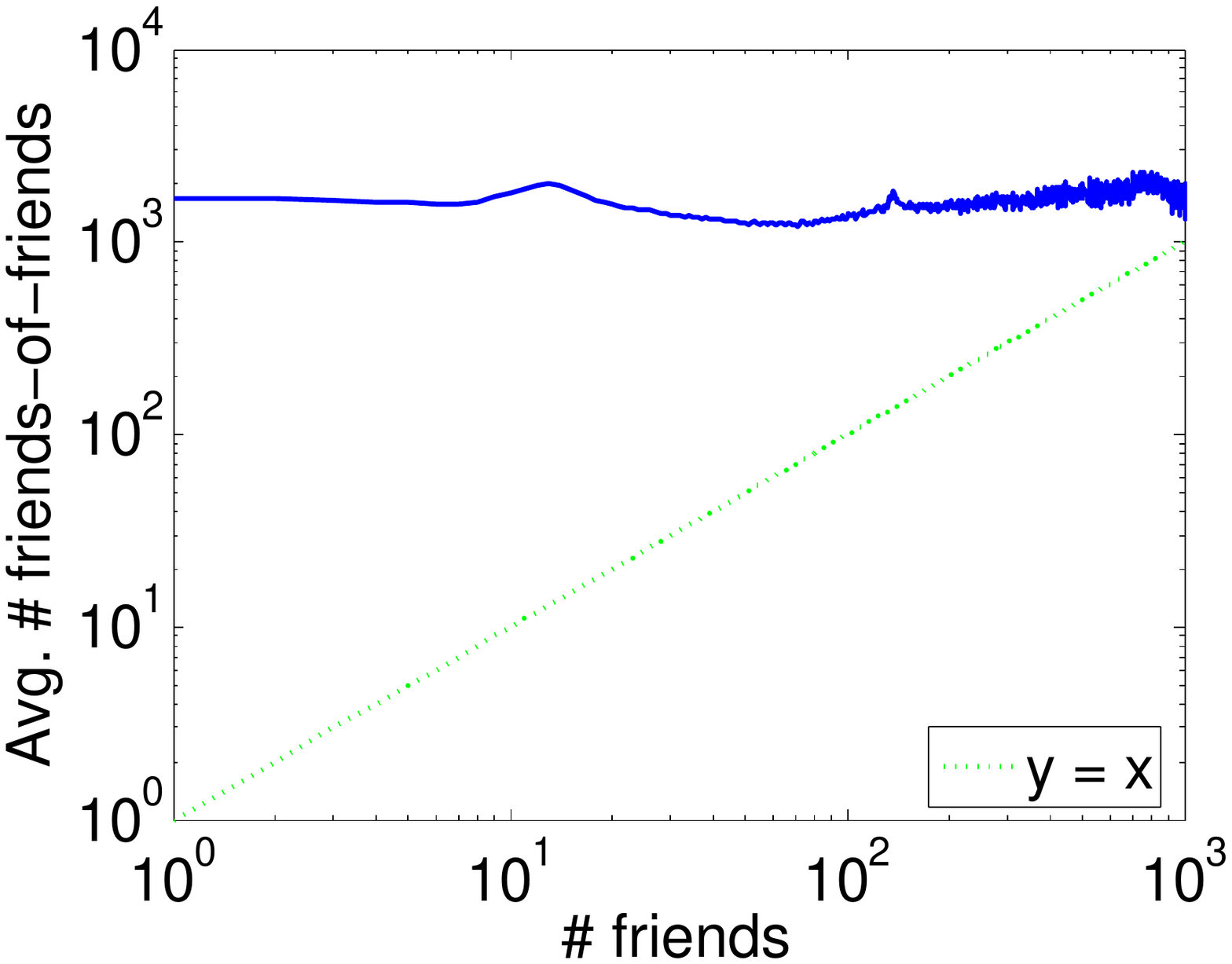}
&
\includegraphics[width=0.5\columnwidth]{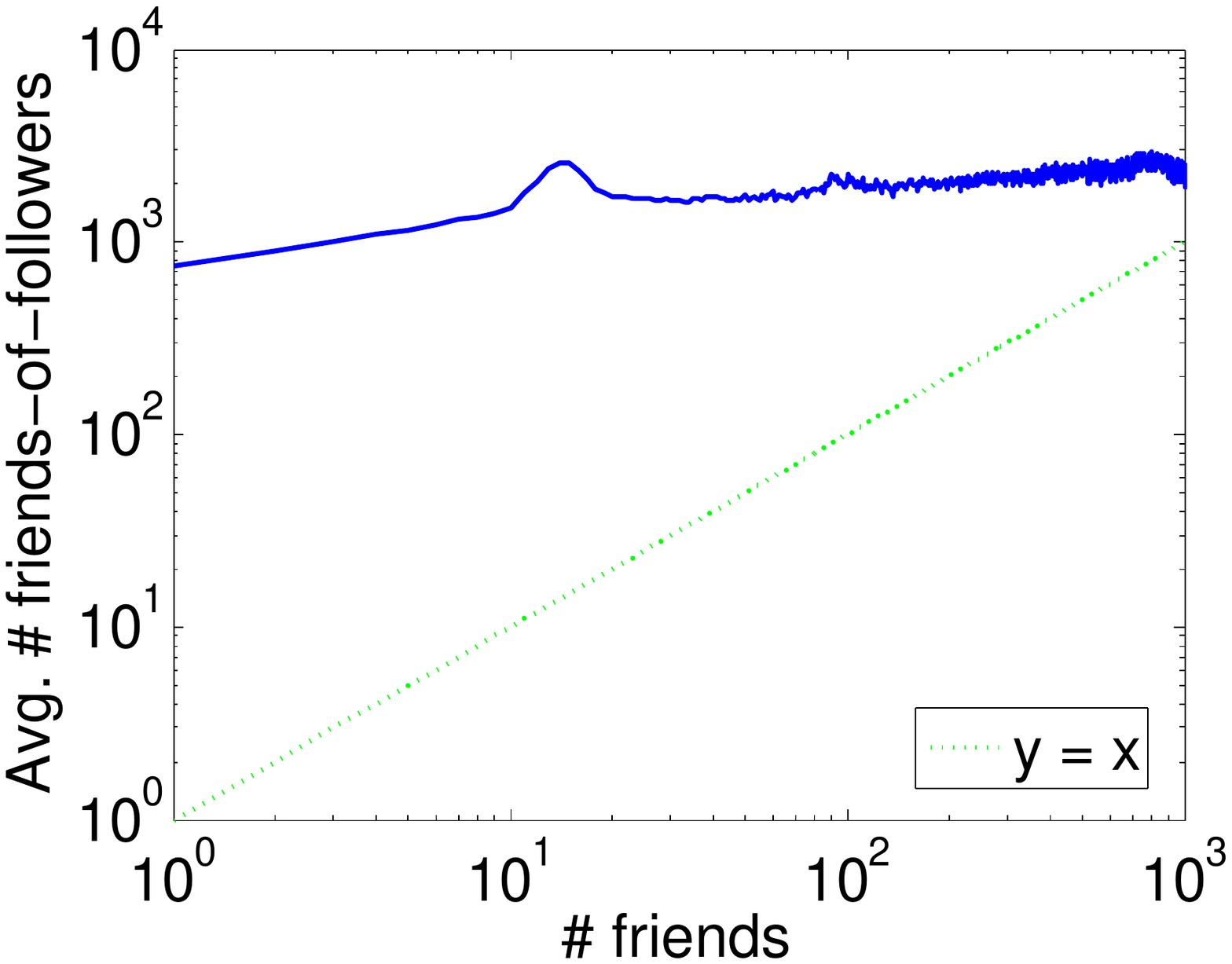}
&
\includegraphics[width=0.5\columnwidth]{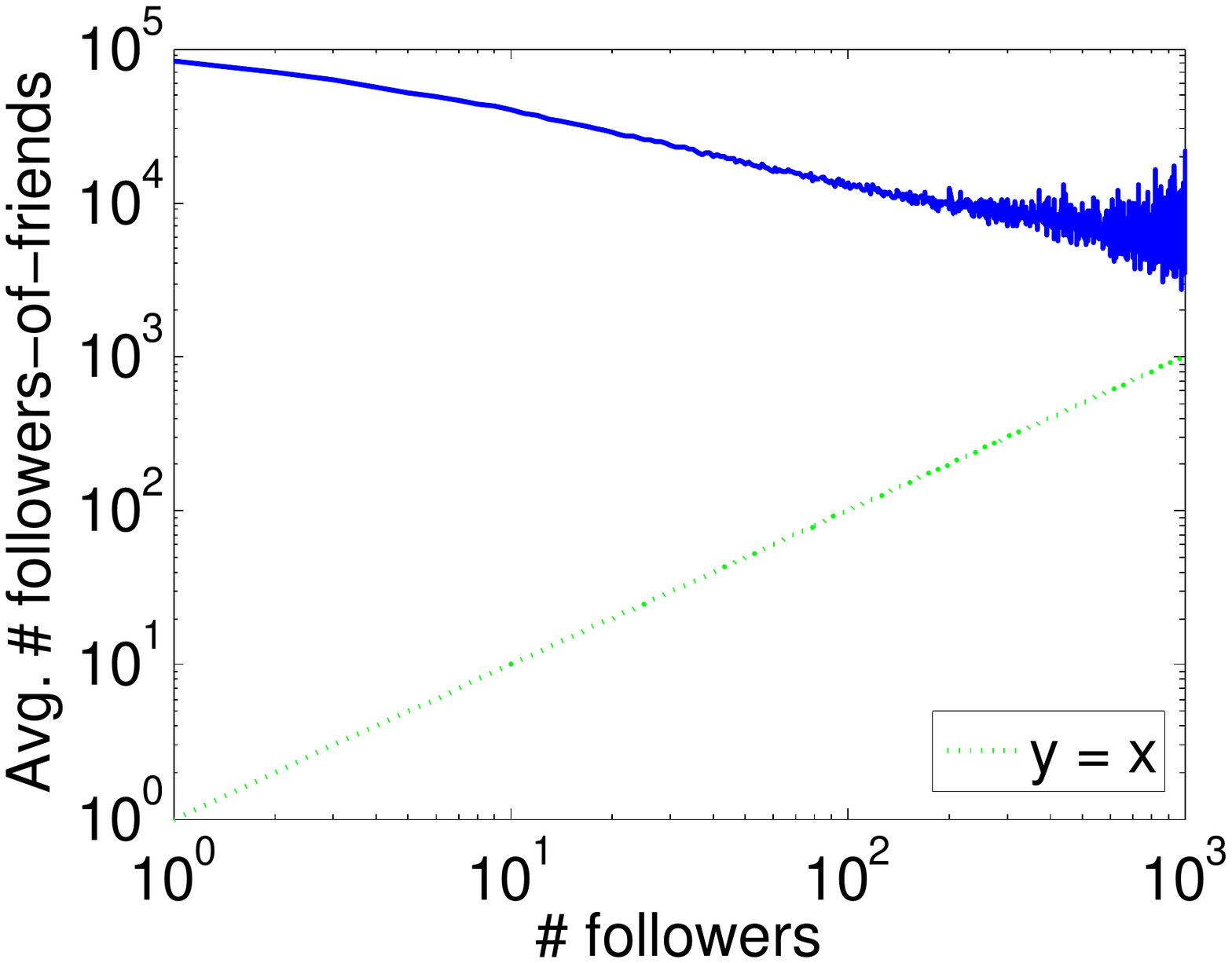}
&
\includegraphics[width=0.5\columnwidth]{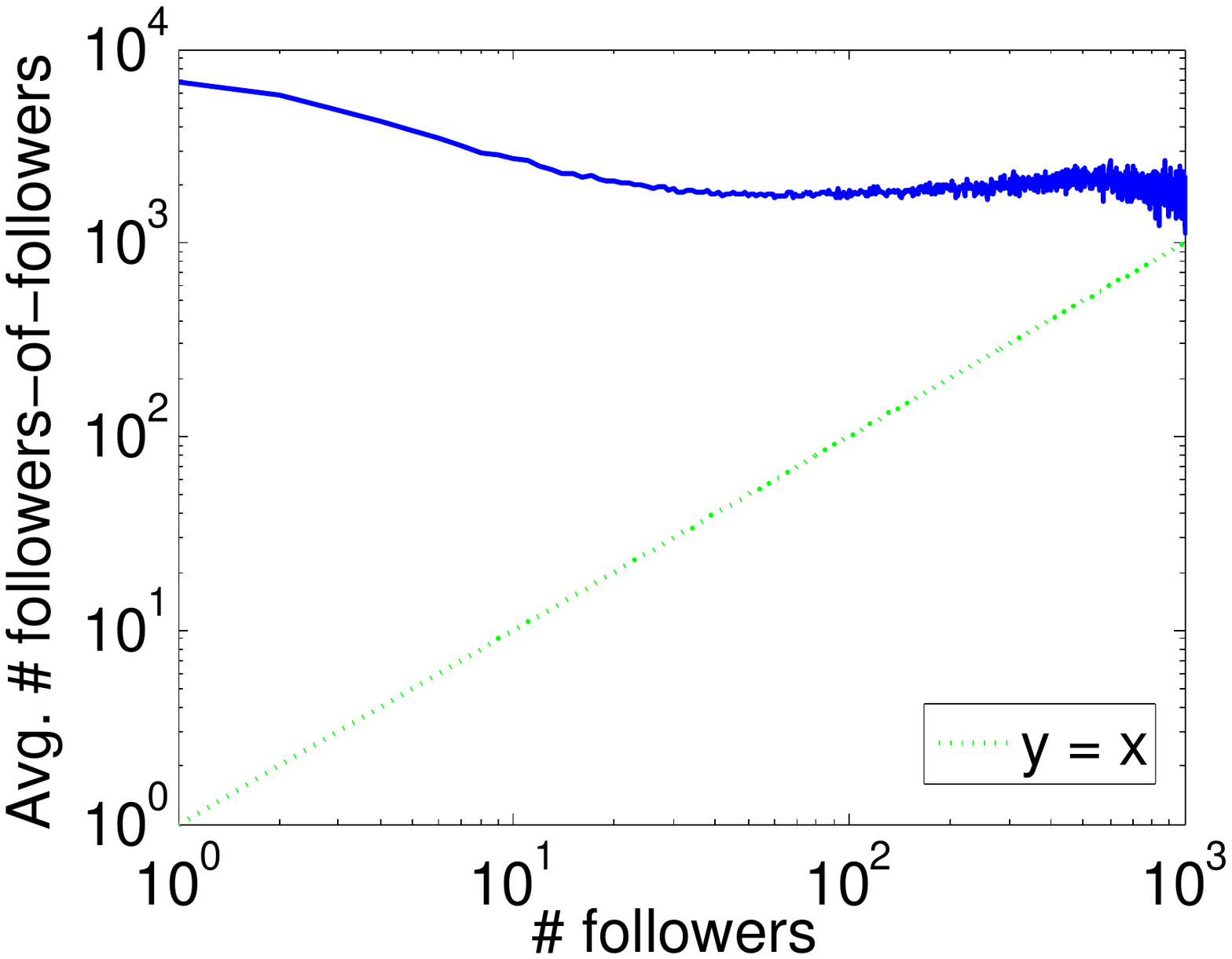}
\\
\includegraphics[width=0.5\columnwidth]{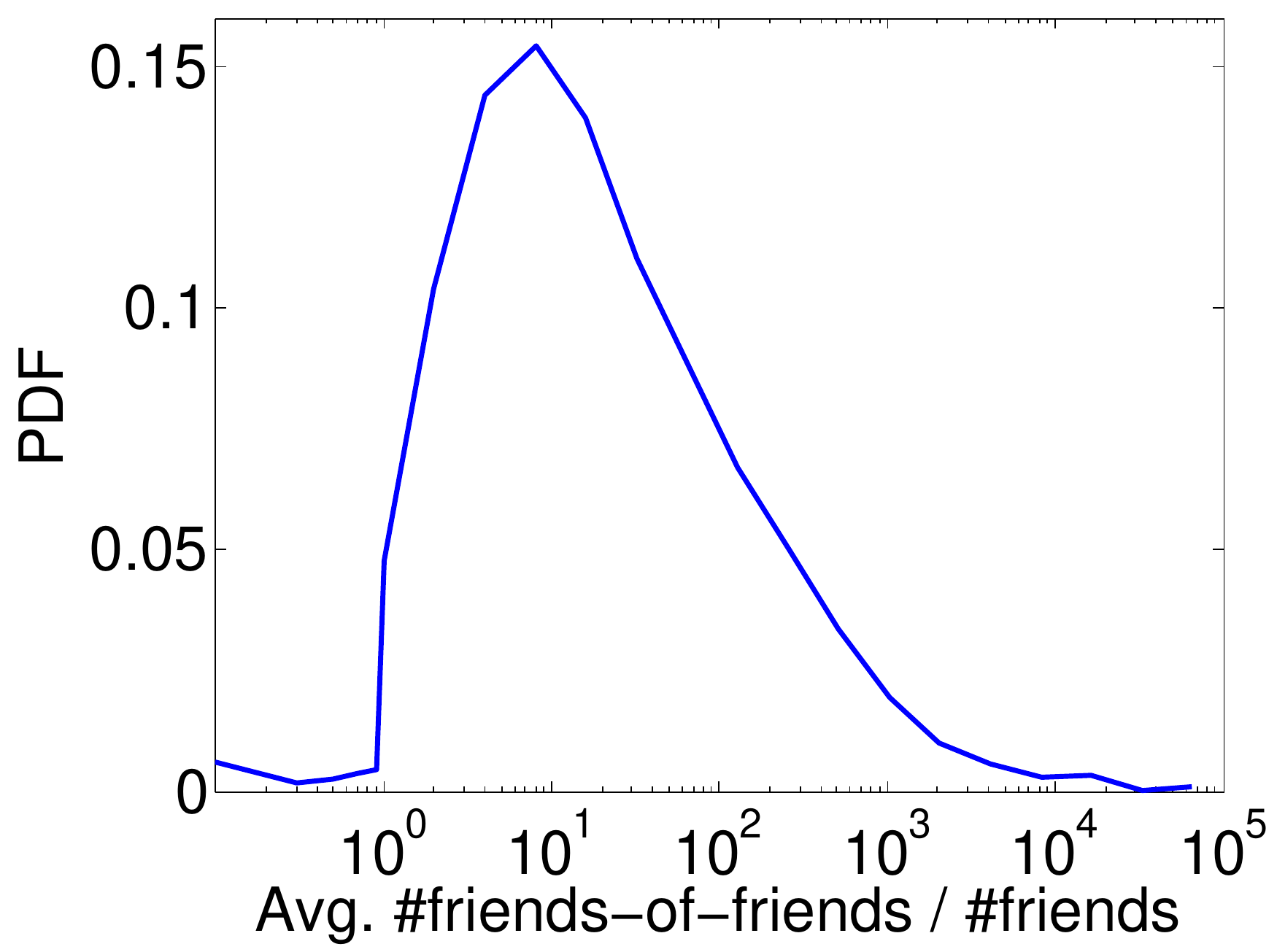}
&
\includegraphics[width=0.5\columnwidth]{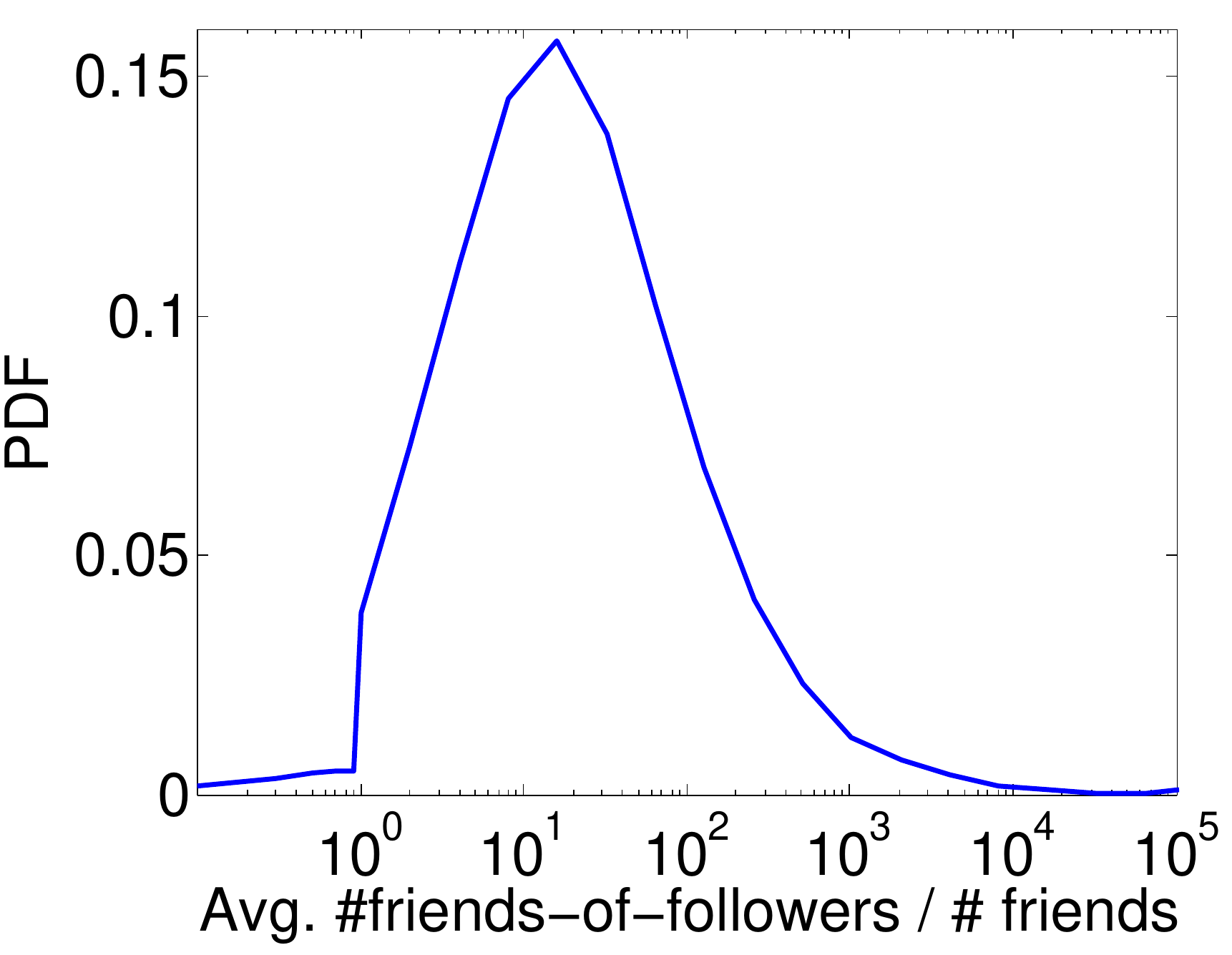}
&
\includegraphics[width=0.5\columnwidth]{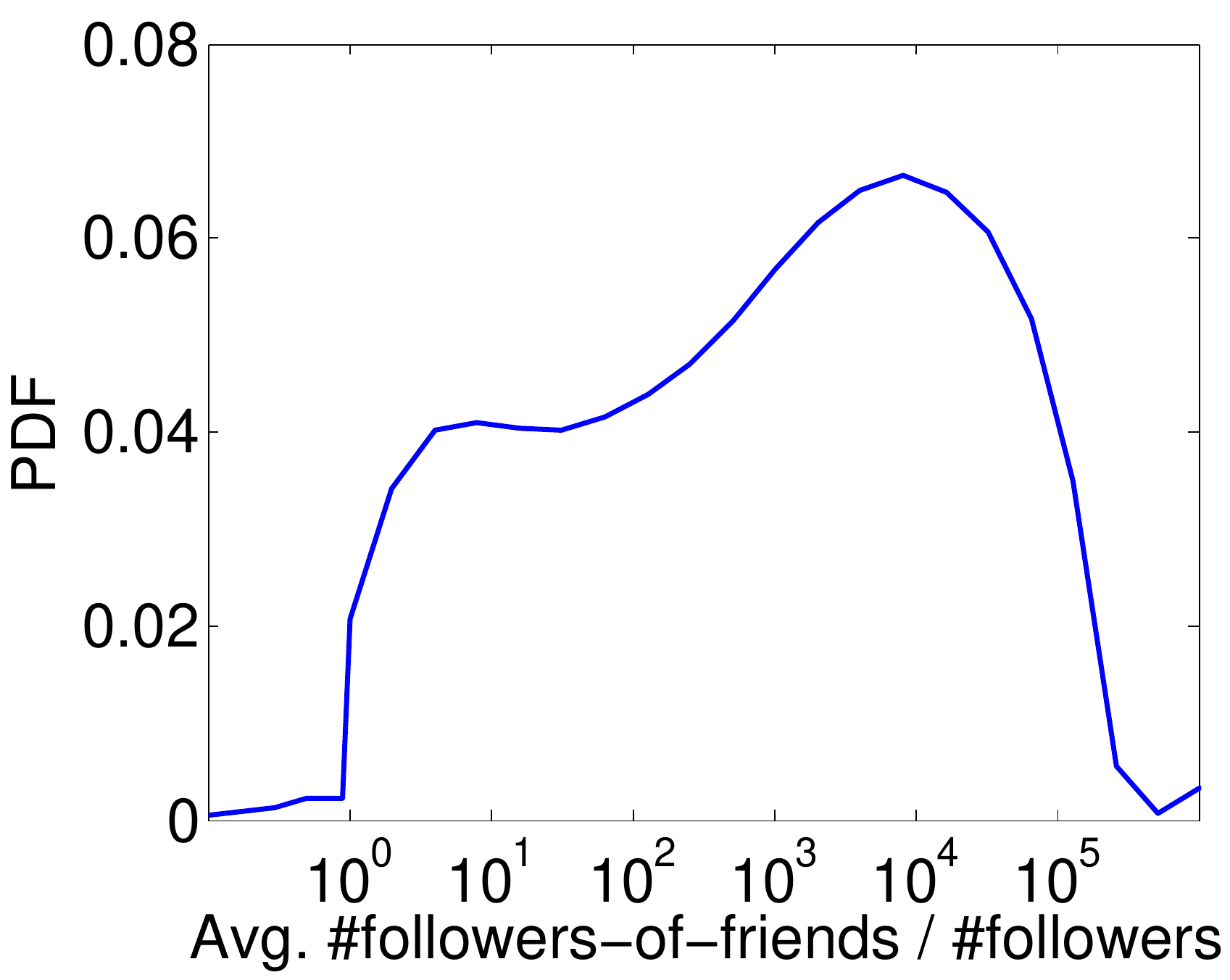}
&
\includegraphics[width=0.5\columnwidth]{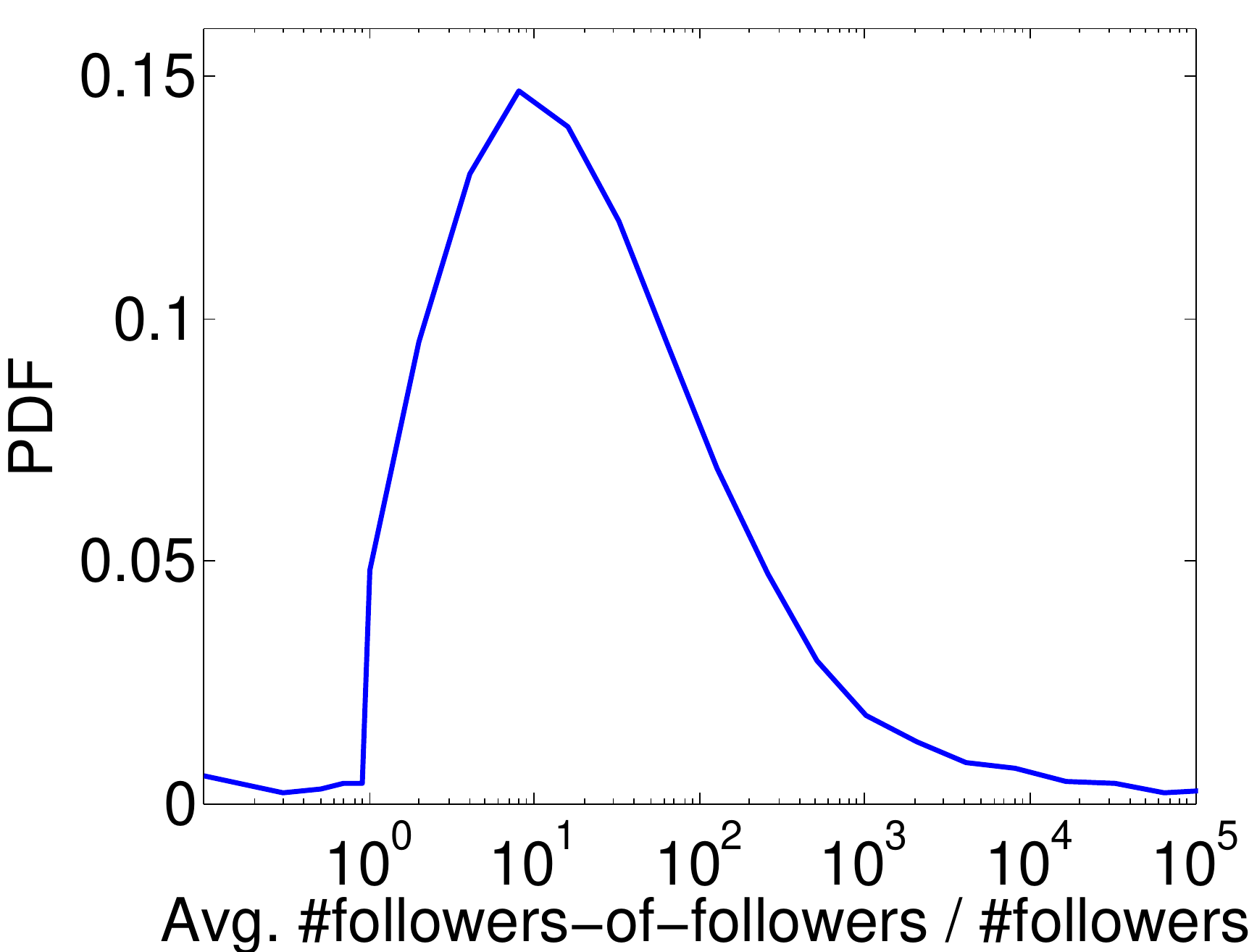}
\\
($i$) & ($ii$) & ($iii$) & ($iv$)
\end{tabular}
\caption{\it Variants of the friend paradox on Twitter showing that your ($i$) friends and ($ii$) followers are better connected than you are (i.e., have more friends on average) and ($iii$, $iv$) are more popular than you are (i.e., have more followers on average). Top row shows the average connectivity (popularity) of user's network neighbors vs user's connectivity (popularity). Data residing above the dashed $y=x$ line indicates "paradox" conditions.  Bottom row shows the probability distribution of the ratio of the average neighbor's connectivity (or popularity) to user's connectivity (or popularity).  Although some users are systematically not in paradox, indicated in the top row, they are a tiny fraction of total users.}
\label{fig:friend_paradox}
\end{center}
\end{figure*}

We empirically validate each statement above. The first statement says that, on average,
a user's friends are better connected than he or she is, i.e., they follow more people
than he or she does. To validate this statement, for each user in the dataset we count
how many friends she has, i.e., how many other users she follows. Then, for each friend,
we count how many other users the friend follows, and average over all friends.
Top Figure~\ref{fig:friend_paradox}($i$) plots the average number of friends-of-friends (ordinate axis)
vs the number of friends (abscissa) a user follows for the users with fewer than 1000 friends. About
99.7\% of users had fewer than 1000 friends. The  line of unit slope shows equality of connectedness. The probability density function (PDF) of the ratio of the average
friend's connectivity to a user's connectivity,  shown in bottom Figure~\ref{fig:friend_paradox}($i$),
is $>1$ for 98\% of the users, peaking around 10. In
other words, in the Twitter follower graph, a typical friend of a user is ten times better
connected than the user.

Not only are a user's friends better connected, but so are the user's followers.
Top Figure~\ref{fig:friend_paradox}($ii$) plots the average number of friends a user's followers
have vs the number of friends the user has for users with fewer than 1000 followers (99.6\% of all users).
Bottom Figure~\ref{fig:friend_paradox}($ii$) shows the PDF of the
ratio of  the friends-of-followers to user's friends. Again, for 98\% of users, this
ratio is above one, indicating that the average follower is better connected than the
user. In fact, a typical follower is almost 20 times better connected than the user is.

The last two variants of the friendship paradox deal with user's popularity, i.e.,
the number of followers he or she has. It appears that on Twitter, user's
both friends and followers are more popular than the user himself of herself. This is shown in
Figures~\ref{fig:friend_paradox}($iii$) and \ref{fig:friend_paradox}($iv$). In our data set,
99\% and 98\% of users were respectively less popular than their friends and followers.
While a typical follower is about 10 times more popular than the user
(Fig.~\ref{fig:friend_paradox}($iv$) bottom),  the ratio of the friend's average popularity to
the user's popularity shows a bimodal distribution (Fig.~\ref{fig:friend_paradox}($iii$) bottom). While some of a user's friends are
ten times more popular, some friends are about 10,000 times more popular, showing a
tendency of Twitter users to follow highly popular celebrities.

\remove{
Figures~\ref{fig:friend_friend} to~\ref{fig:friend_followers} show that the paradox
exists in all the four cases. Figure~\ref{fig:friend_friend_ave} shows the average
number of friends of friends averaged over all users with the same number of friends.
Average number of friends of friends is shown for the users with less than 1000
friends (ZZ\% of users). The figure also shows the line {\it y = x}, which
means friends of the user have the same number of friends as the user herself on
average and values above the line correspond to the fact that the friends have
more friends on average. Moreover, Figure~\ref{fig:friend_friend_pdf} shows the
probability distribution function (PDF) of ration of average number of friends of
friends over number of friends. The figure show that the vast majority of users
(98\%) have ratio larger than one, which means that friends of 98\% of users have
more friends on average compared to themselves. A similar situation exists for
other three cases of friendship paradox on a directed graph
(Figures~\ref{fig:follower_friend},~\ref{fig:follower_followers}, and~\ref{fig:friend_followers}).
}

\subsection{Friend Activity Paradox}
In addition to connectivity and popularity paradoxes, we also demonstrate a novel activity paradox on Twitter.

\begin{quotation} \emph{Friend activity paradox:  On average, your friends are more active than you are.}
\end{quotation}

\begin{figure}[tb]
\begin{center}
\subfigure[Average number of tweets posted by user's friends vs the number of tweets posted by the user. ]{%
\label{fig:friend_tweet_ave}
\includegraphics[width=0.4\textwidth]{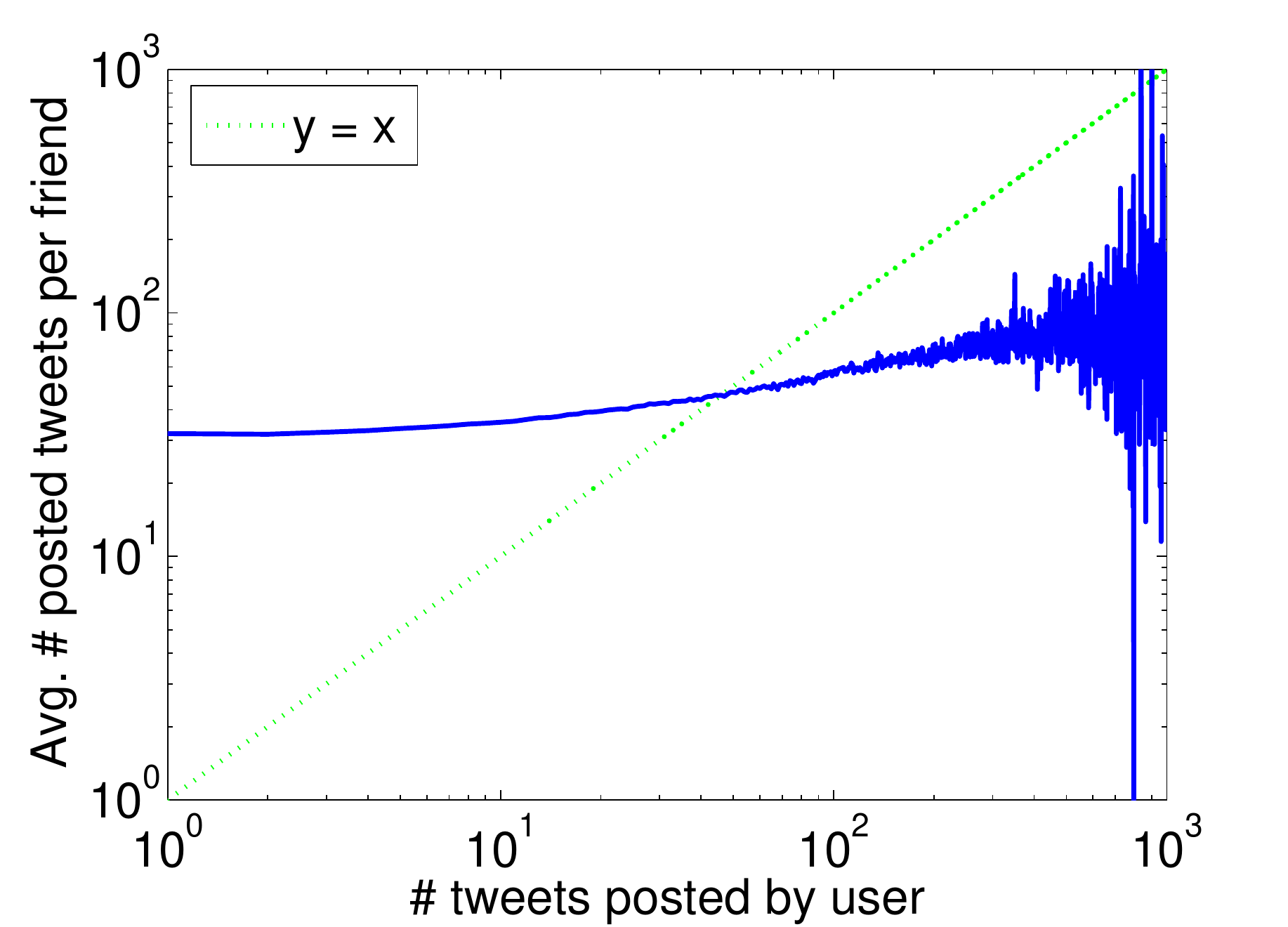}
}\\
\subfigure[PDF of the ratio of tweets posted by friends and tweets posted by number of posted tweets.]{%
\label{fig:friend_tweet_pdf}
\includegraphics[width=0.4\textwidth]{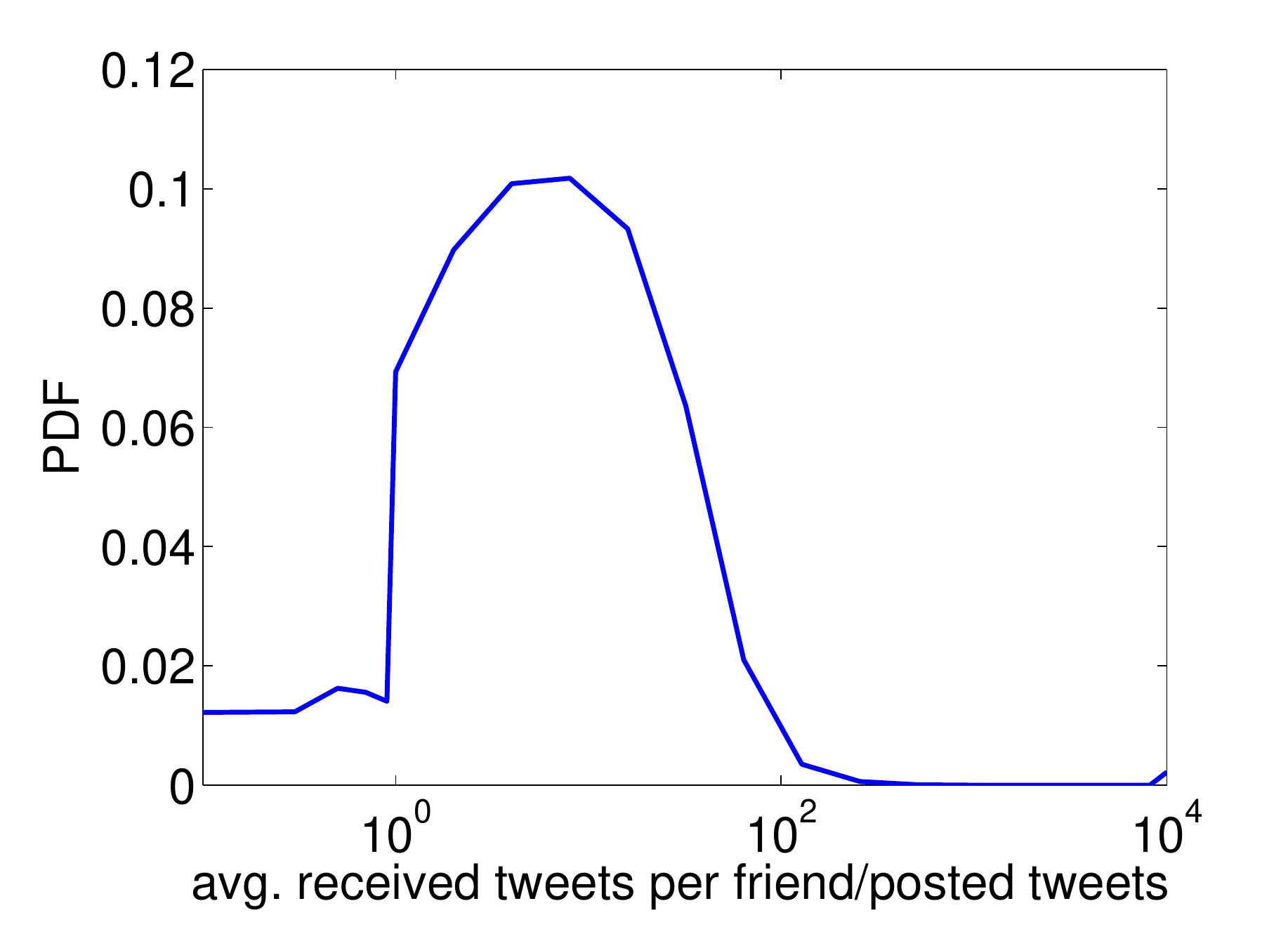}
}
\caption{\it Comparison of user's activity and the average activity of his or her friends (measured by the number of tweets posted by them). Most (88\%) of the users are less active than their friends on average.}
\label{fig:friend_tweet}
\end{center}
\end{figure}

To empirically validate this paradox, we measure user activity, i.e., the number of
tweets posted by a user during a given time period; we exclude users who joined Twitter after the start of
the time period. After windowing by a two-months time period we are left with 37M tweets
from 3.4M users and 144.5M links among these users. Note that the dataset contains a random
sample of all tweets; therefore, the number of tweets posted by
the user in our sample is an unbiased measure of his or her overall activity.
At the same time, we measure the number of sampled tweets posted by user's friends during the
same time interval. Figure~\ref{fig:friend_tweet_ave}  shows the average activity (number
of posted tweets ) per friend of users who each have same level of activity, i.e.,
mean average friend activity as a function of user activity. The unit slope $y=x$ line is shown for
comparison.
88\% of all users are less active than their typical friend. Figure~\ref{fig:friend_tweet_pdf} shows the probability
distribution of the ratio of average per friend activity over user activity. For the vast
majority of users, the friend activity paradox holds: their friends are more active than they are.

It is known that some users become inactive after some time. To ensure that our results
are not affected by inactive users, we checked the same paradox for a shorter time
period of one week, during which time fewer users may have become inactive. Activity paradox still holds. In fact, a much larger
fraction of users are in the paradox regime: 99\% of users are less active than their friends.
Also, note that in all the analyses that we are comparing users with their friends (followers)
we exclude users who don't have any friends (followers), because there is no one for the comparison.

\subsection{Virality Paradox}
Your friends' superior social connectivity puts them  in a better position to monitor, in aggregate, the flow of information, thereby mediating the information you receive via the social network. Perhaps this also puts them in a position to receive higher quality content.  As a measure of quality, we investigate virality
of URLs tweeted by users, i.e., number of times a URL was posted by any user over some time period.

\begin{quotation} \emph{Virality paradox: On average, your friends spread more highly viral content than you do.}\end{quotation}

\begin{figure}[tb]
\begin{center}
\subfigure[PDF of  $\langle$size of posted cascade per friend$\rangle$~/~$\langle$size of posted cascades$\rangle$.]{%
\label{fig:post_friend_size_pdf}
\includegraphics[width=0.4\textwidth]{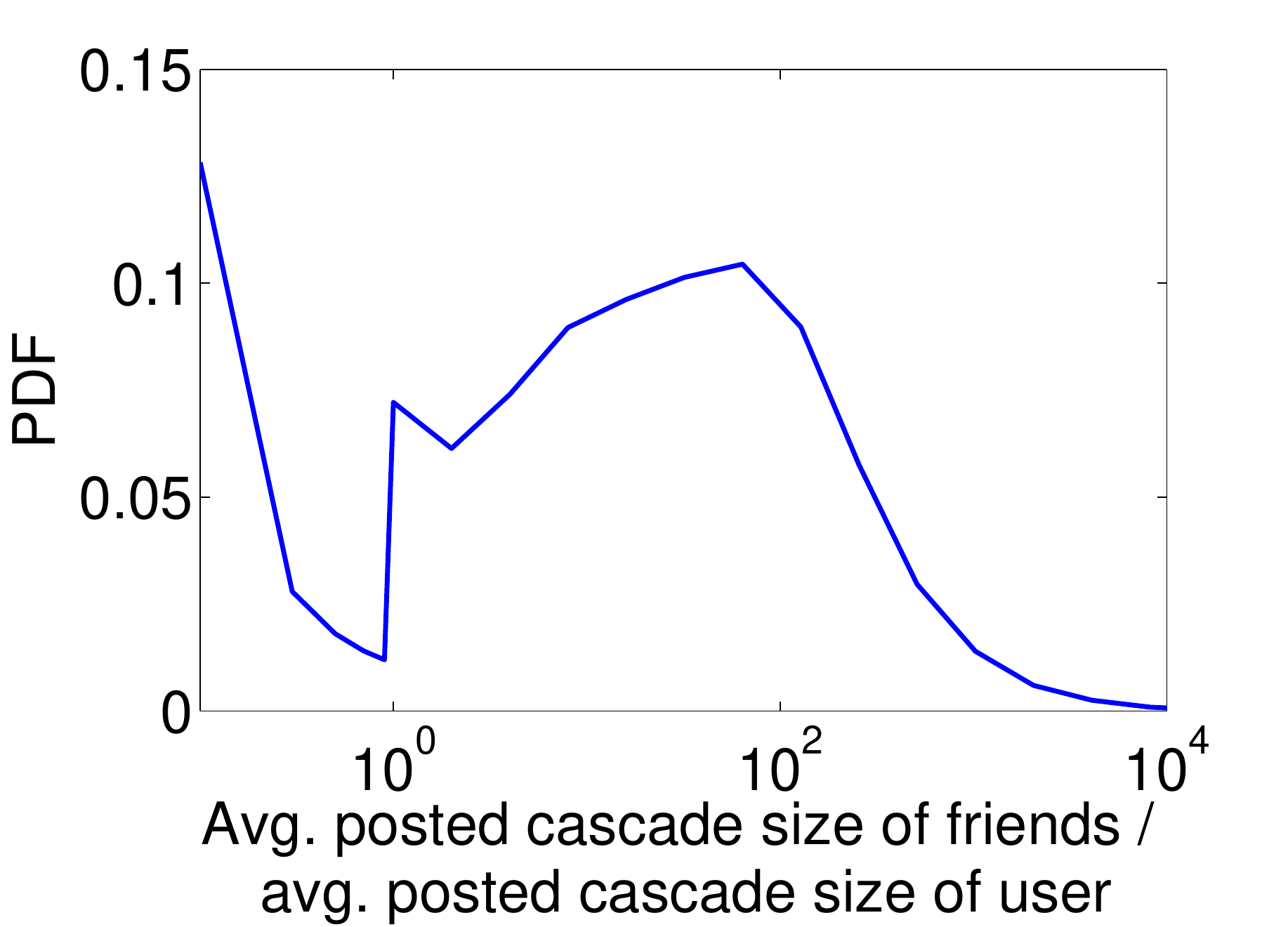}
}\\
\subfigure[PDF of  $\langle$size of received cascades per friend$\rangle$~/ $\langle$size of received cascades$\rangle$.]{%
\label{fig:rec_friend_size_pdf}
\includegraphics[width=0.4\textwidth]{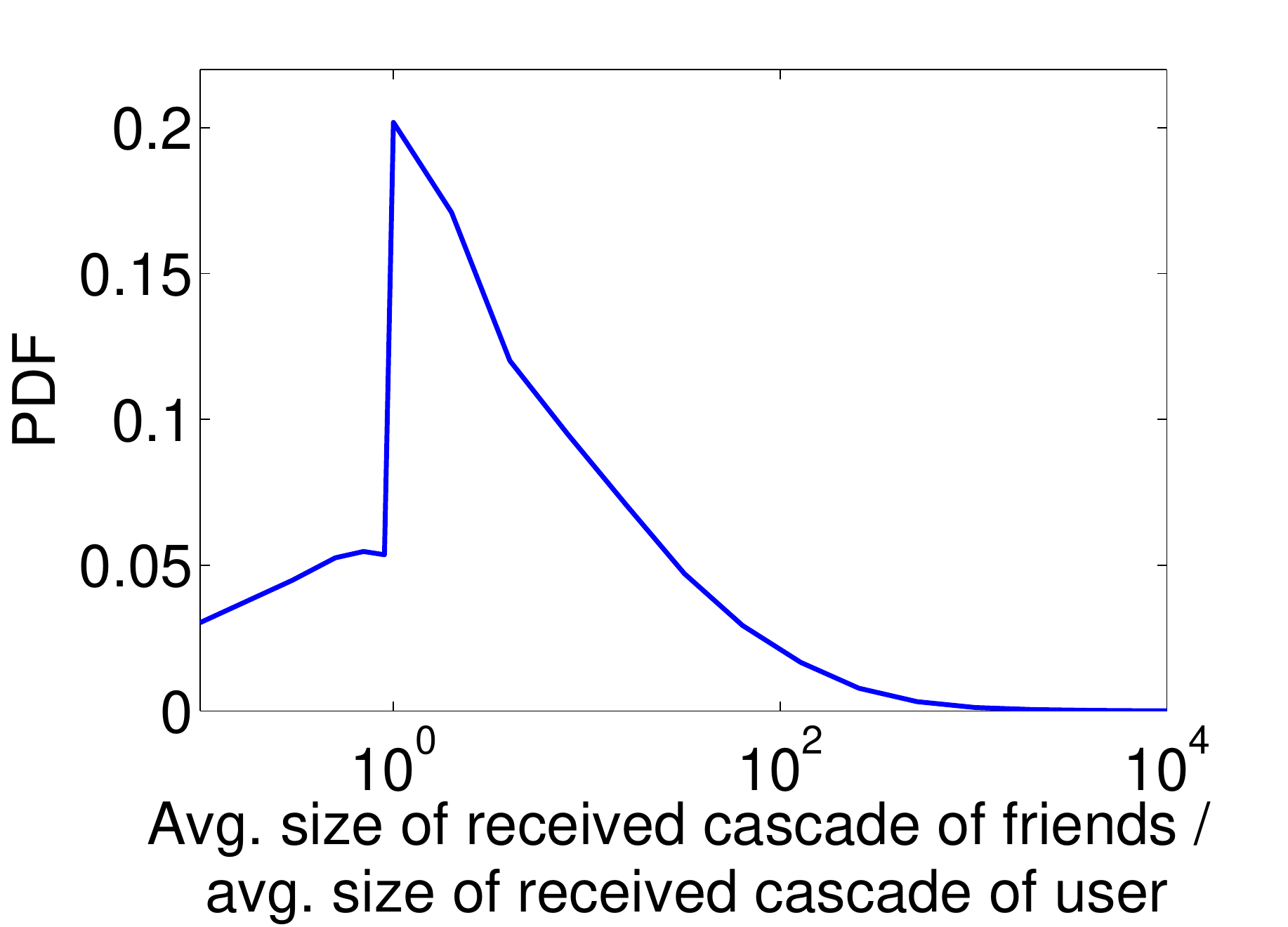}
}
\caption{\it Comparison of average size of posted and received cascade of users with their friends. For the vast majority of users, their friends both receive and post URLs with higher average cascade size, indicating a virality paradox.}
\label{fig:activity2}
\end{center}
\end{figure}

To confirm this paradox, we calculate average size of posted URL cascades for each user and
compare this value with the average size of posted cascades of friends. We observe that $32\%$ of users
haven't posted any URLs (average cascade size of $0$), while their friends did. Therefore,
these inactive users have posted fewer viral cascades than their friends. For the remaining $68\%$
of users, Figure~\ref{fig:post_friend_size_pdf} shows the probability distribution of the ratio of average size of
cascades posted by friends to the average size of cascades posted by user. We find that $79\%$ of users have
ratio of greater than 1, which means that their friends have posted more viral content. Considering
the users who haven't post any URLs, $86\%$ of all users have posted less viral content than their friends.

Users not only post less popular URLs than their friends, but also receive less viral content
than their friends do, on average. Figure~\ref{fig:rec_friend_size_pdf} shows the probability distribution of the ratio of the average size of cascades friends receive to the average size of cascades received by the user. Here again $76\%$ of users receive smaller (less viral) cascades than their friends ($15\%$ of users have received URLs with same level of virality as their friends).

\subsection{Spam Filtering}
One trivial explanation of our results could be the presence of spammers in our sample.
Spammers generate more tweets than normal users, so their presence in our sample could bias
our estimates of user activity.

To validate that the paradoxes don't exist because of the spammers, we eliminated spammers
from the dataset in two different ways. First, we use the set of spammers from~\cite{www2012ghosh}.
These users' profile was suspended by Twitter authorities and also the users posted at least one
blacklisted URL. Second, we took the approach of~\cite{Ghosh11snakdd} and classified users
as spammers based on entropy of content generated and entropy of time intervals
between tweets (spammers tend to have low entropy of content and tweeting time intervals).

In both cases, after removing all spammers from the network and excluding their tweets,
all paradoxes still hold. In fact, in some cases the paradox becomes even stronger. For example,
if we exclude users based on their content and activity entropy, 93\% of users would be less
active than their friends (instead 88\% before spam filtering).

\begin{figure}[tb]
\begin{center}
\subfigure[Average number of tweets received by users with the given number of friends] {%
\label{fig:friend_rec}
\includegraphics[width=0.4\textwidth]{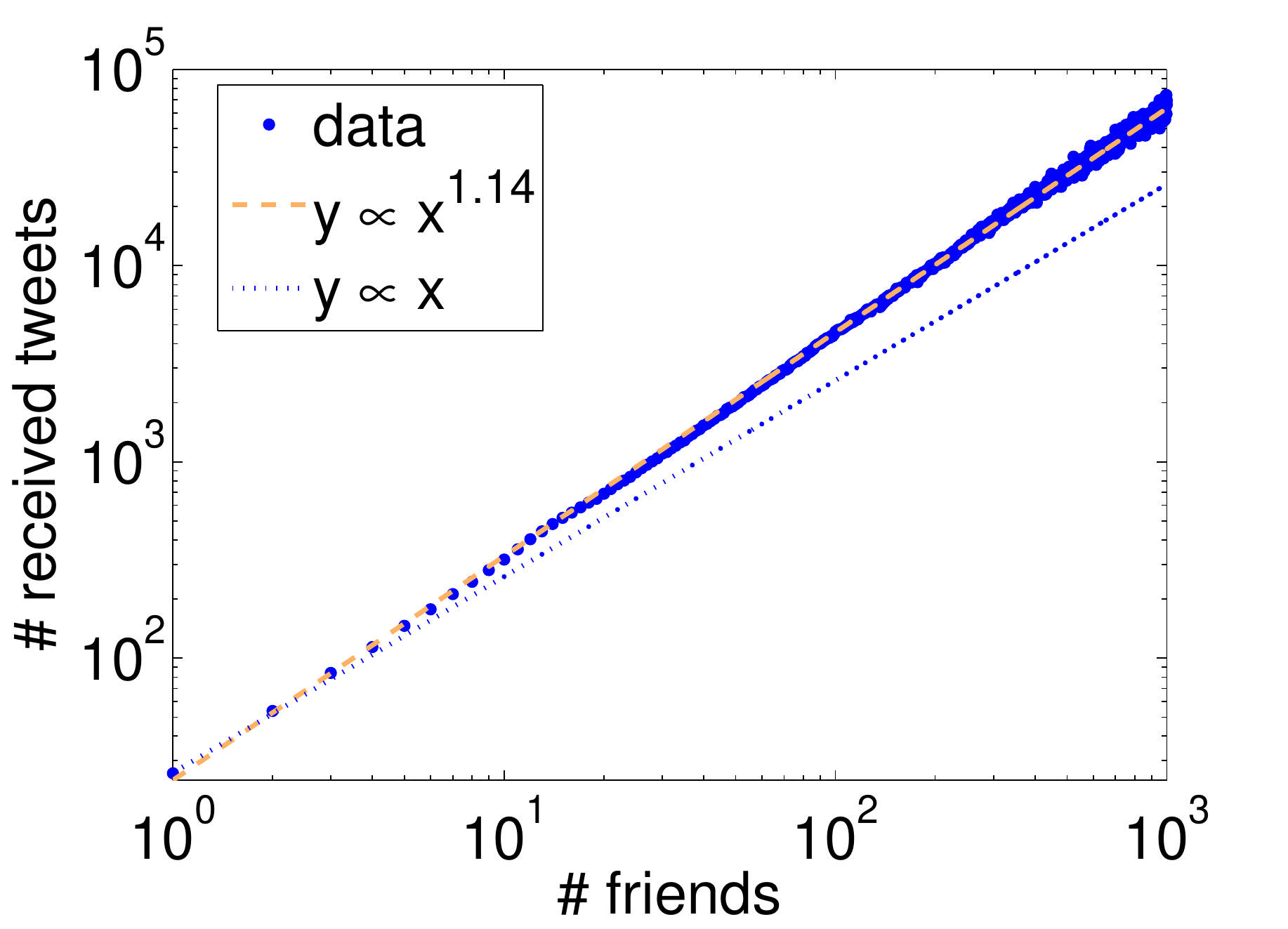} 
} \\
\subfigure[Average number of tweets posted by user vs the number of received tweets] {
\label{fig:rec_tweet}
\includegraphics[width=0.4\textwidth]{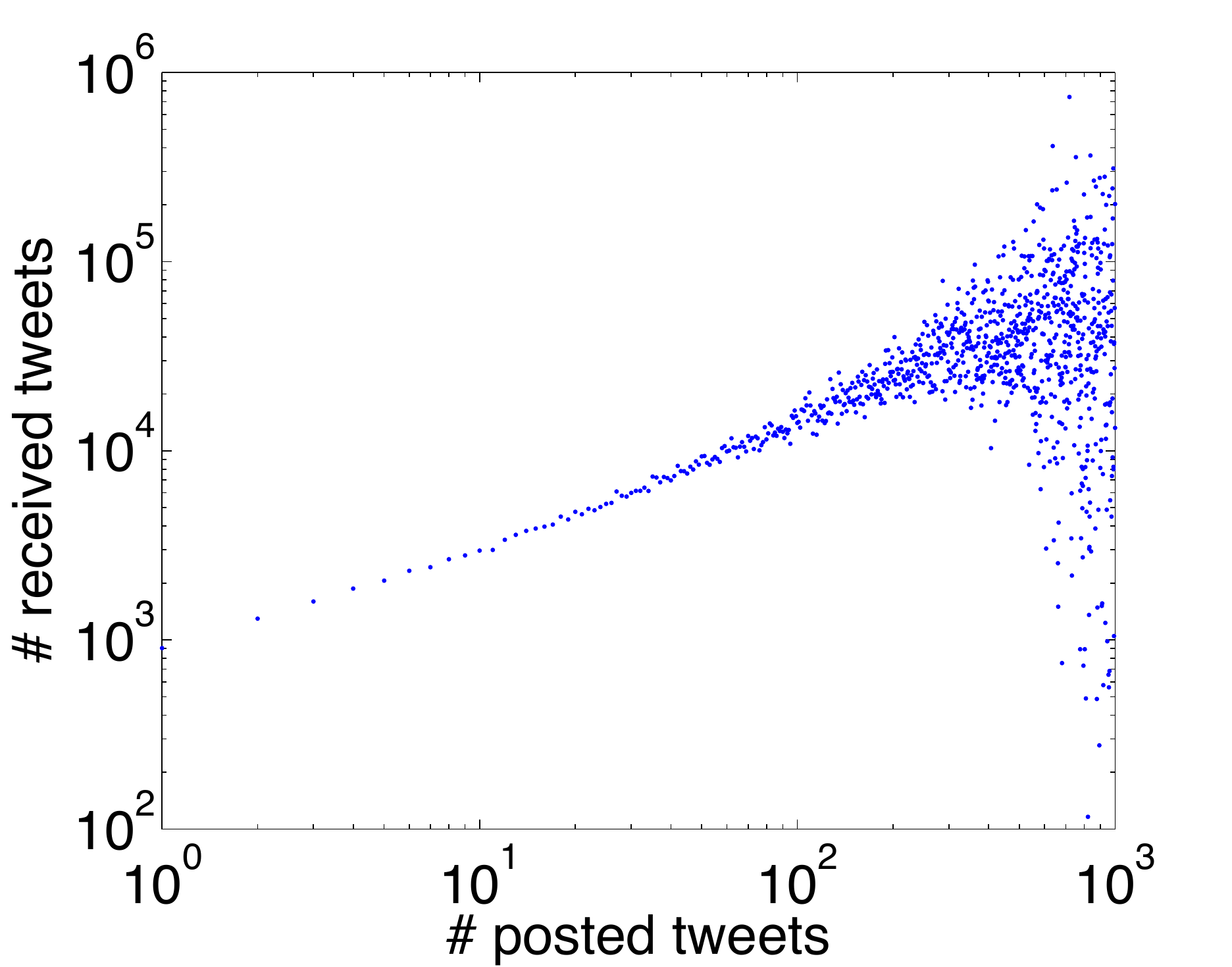}
}
\caption{\it Growth in the volume of incoming information as a function of user's connectivity and user activity it stimulates. Lines in (a) show the best power law and linear fits.}
\label{fig:activity}
\end{center}
\end{figure}

\begin{figure}[tb]
\begin{center}
\subfigure[Average number of posted tweets vs number of followers.]{%
\label{fig:follower_tweet}
\includegraphics[width=0.4\textwidth]{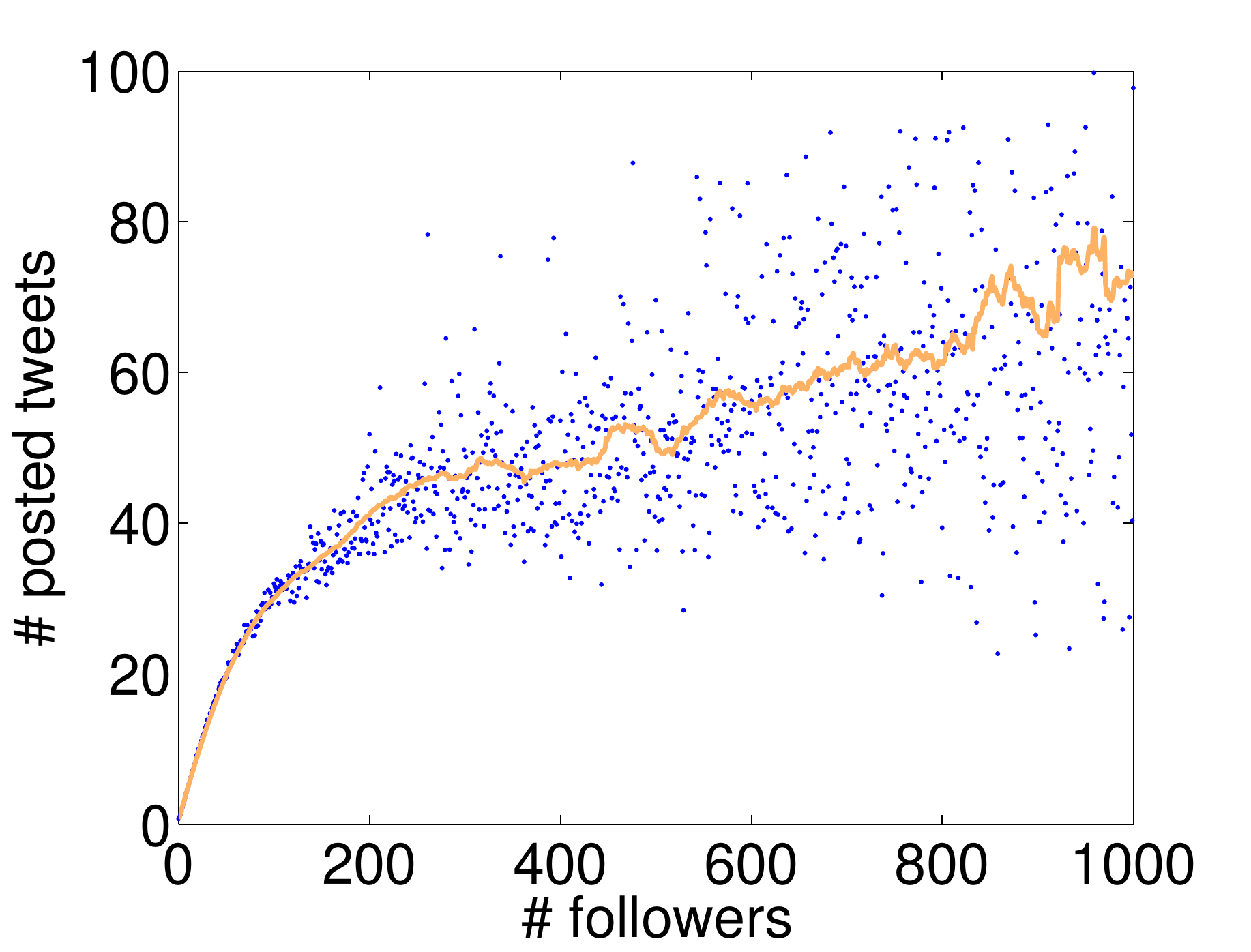}
}\\
\subfigure[Average number of posted tweets vs number of friends. ]{%
\label{fig:friend_tweet3}
\includegraphics[width=0.4\textwidth]{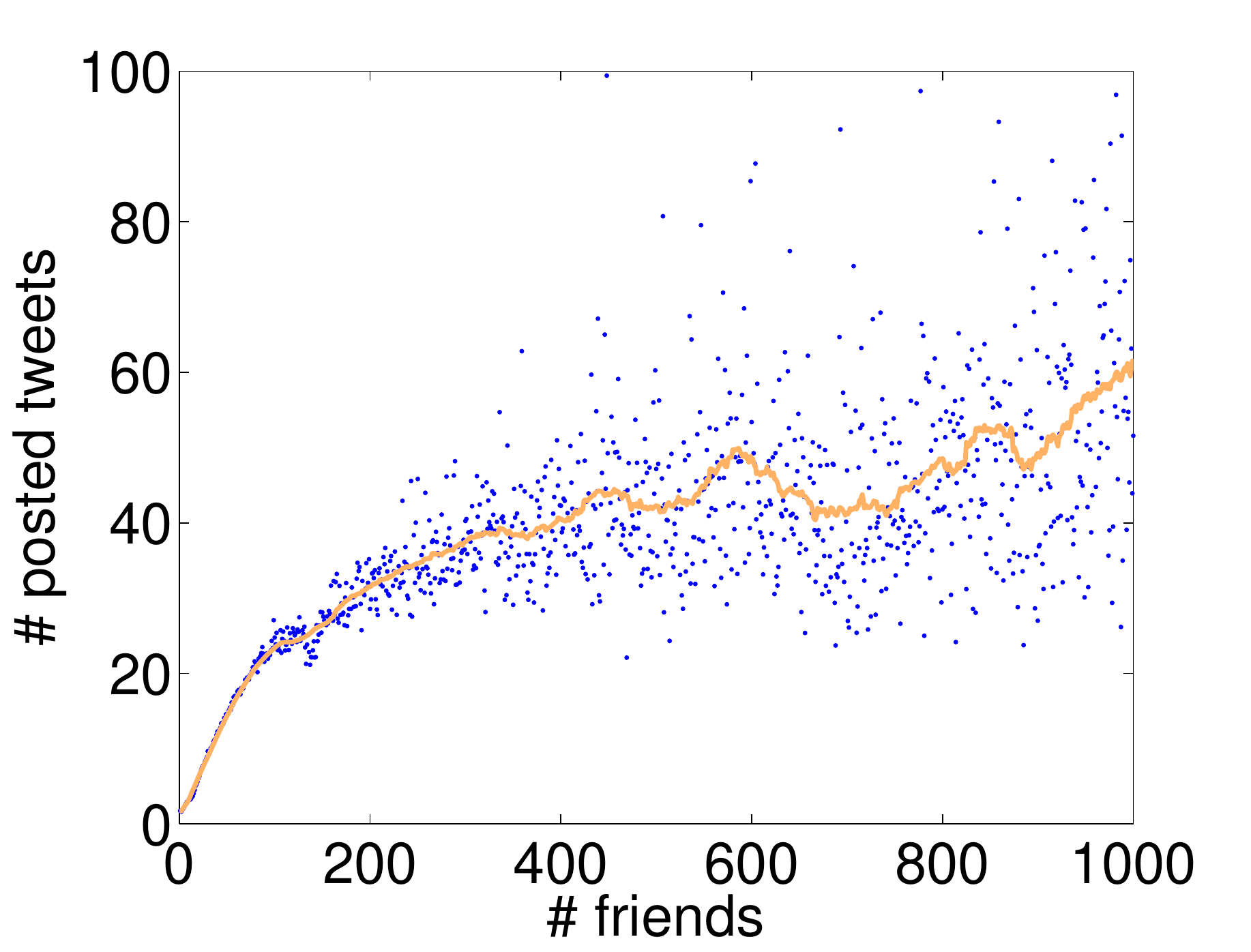}
}
\caption{\it User activity as a function of the number of followers and friends the user has.}
\label{fig:activity3}
\end{center}
\end{figure}

\section{Friend Paradox and Information Overload} \label{sec:cognitiveload}

\noindent The friend activity paradox in directed social networks of online social media is not a mere statistical curiosity --- it has surprising implications for how social media users process information. As  social media users become more active on the site, they may want to grow their social networks to receive more novel information. Clearly, adding more friends will increase the amount of information a user has to process. 
However, according to the friend activity paradox, an average new friend is more active than the user is herself; therefore, the volume of new information in a user's stream will grow super-linearly as new connections are added. Sometimes the volume of new information will exceed user's ability to process it, pushing the user into information overload regime. Overloaded users are less sensitive detectors of information.

\subsection{User Activity and Incoming Information Volume}

\noindent We study how the \emph{volume of incoming information}, measured by the number of tweets received by a user, grows with the size of a user's social network. Figure~\ref{fig:friend_rec} shows
the average number of tweets received by users who follow a given number of friends. The data is shown for users with up to 2000 friends, and has surprisingly low dispersion. This data is best fit  by an power-law function with exponent 1.14 ($R^2=0.9865$). The best linear fit has slope of 71 ($R^2 = 0.8915$), while the best quadratic fit has slope of 60  ($R^2=0.8930$). The lines in Figure~\ref{fig:friend_rec} show the best power-law and linear fits, where the linear fit was shifted down vertically for clarity. These data show that  the average volume of information received by a user grows super-linearly with the number of friends! Regardless of the precise functional form, the volume of incoming information increases quickly with user's connectivity: for every new friend, users receive hundreds of new posts in their stream.\footnote{This total is over the course of two months. Our dataset is a 20\% sample, so the total numbers should be scaled accordingly.}

Users can compensate for the increased volume of incoming information by increasing their
own activity, e.g., visiting Twitter more frequently. While we cannot directly observe
when a user visits Twitter to read friends' posts, we can indirectly estimate user
activity by counting the number of tweets he or she posts within the time period.
Figure~\ref{fig:rec_tweet} shows that users who receive more information are also more
active, though after about 500 posted tweets (over a two month period) the relationship
between incoming volume of information and user activity becomes very noisy. These
extremely active users (posting 50 or more tweets a day, on average, accounting for our 20\% sample), who are not
limiting how much information they receive, could be spammers. We include them, because their activity impacts the information load of people who choose to follow them.

\begin{figure*}[tbh]
\begin{center}
\begin{tabular}{@{}c@{}c@{}c@{}c@{}}
\includegraphics[width=0.5\columnwidth]{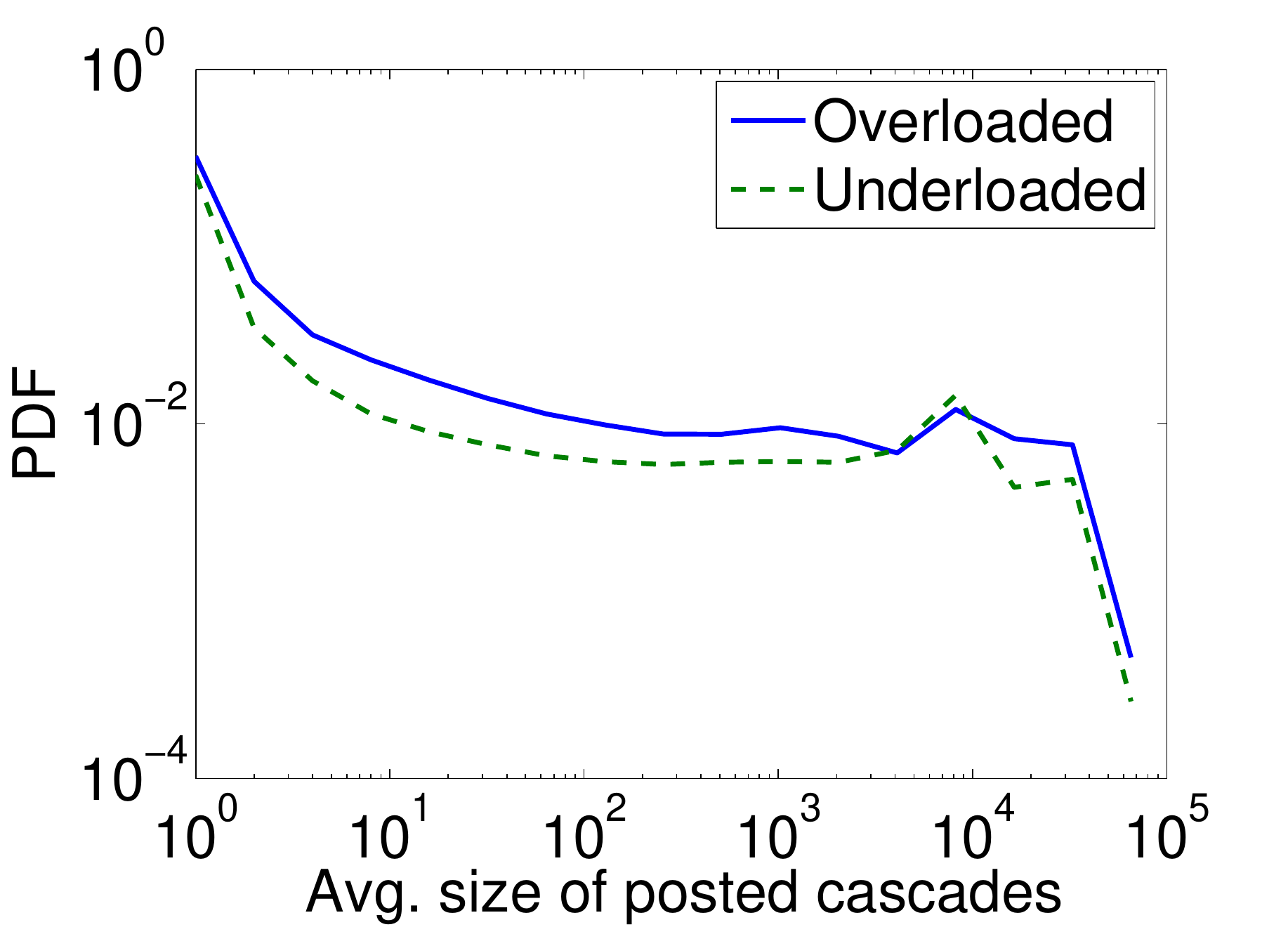}
&
\includegraphics[width=0.5\columnwidth]{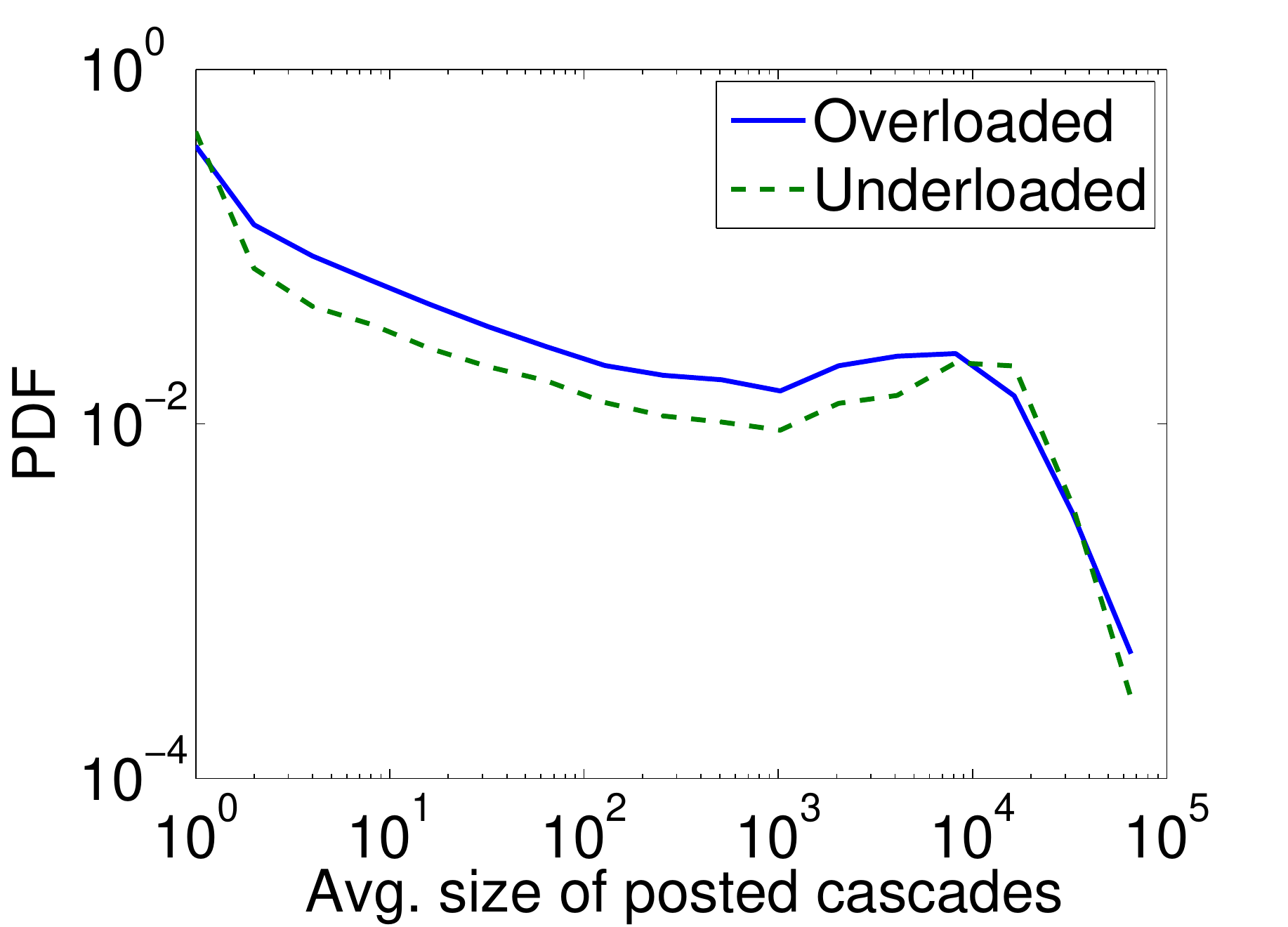}
&
\includegraphics[width=0.5\columnwidth]{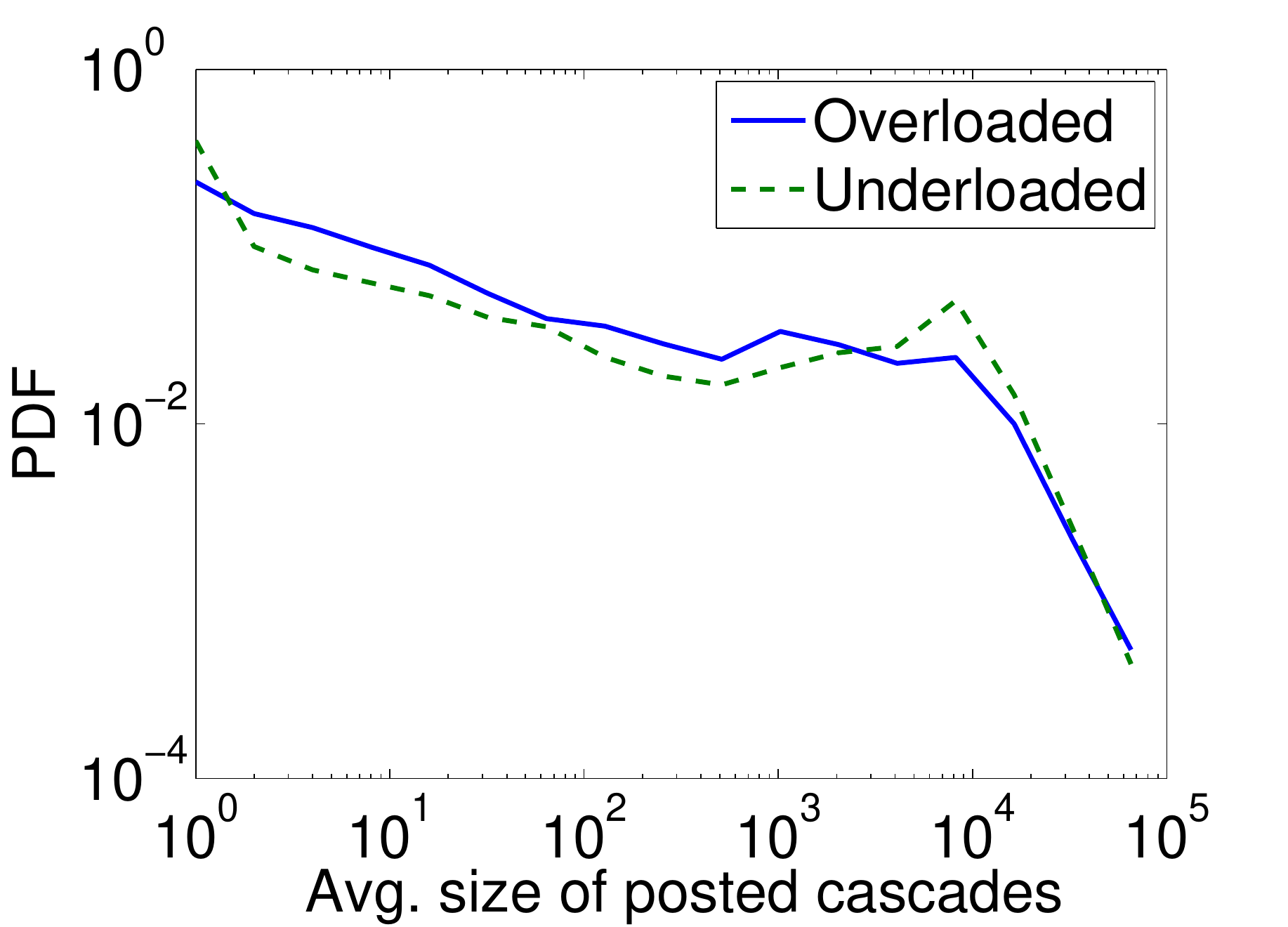}
&
\includegraphics[width=0.5\columnwidth]{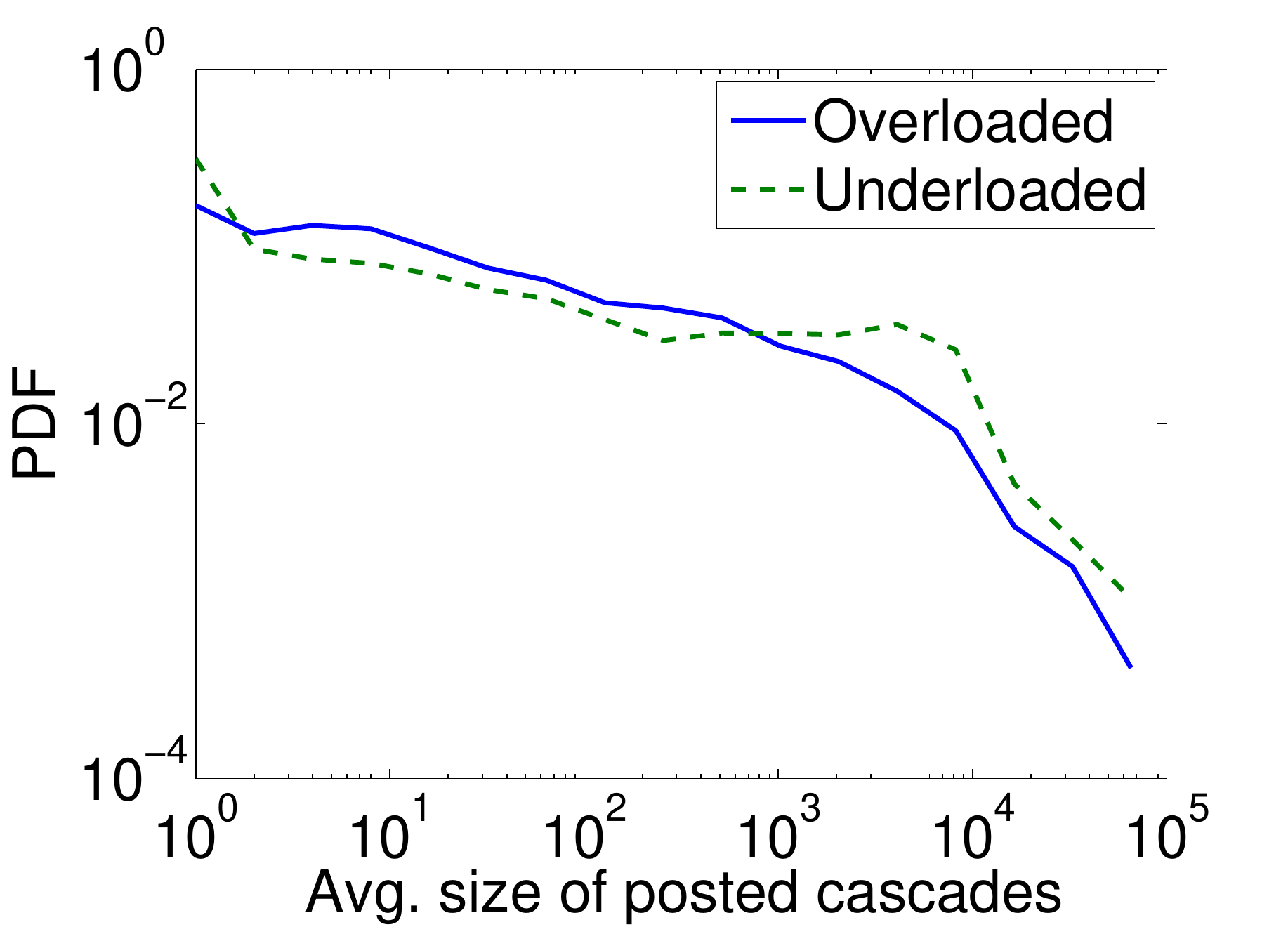}
\\
\includegraphics[width=0.5\columnwidth]{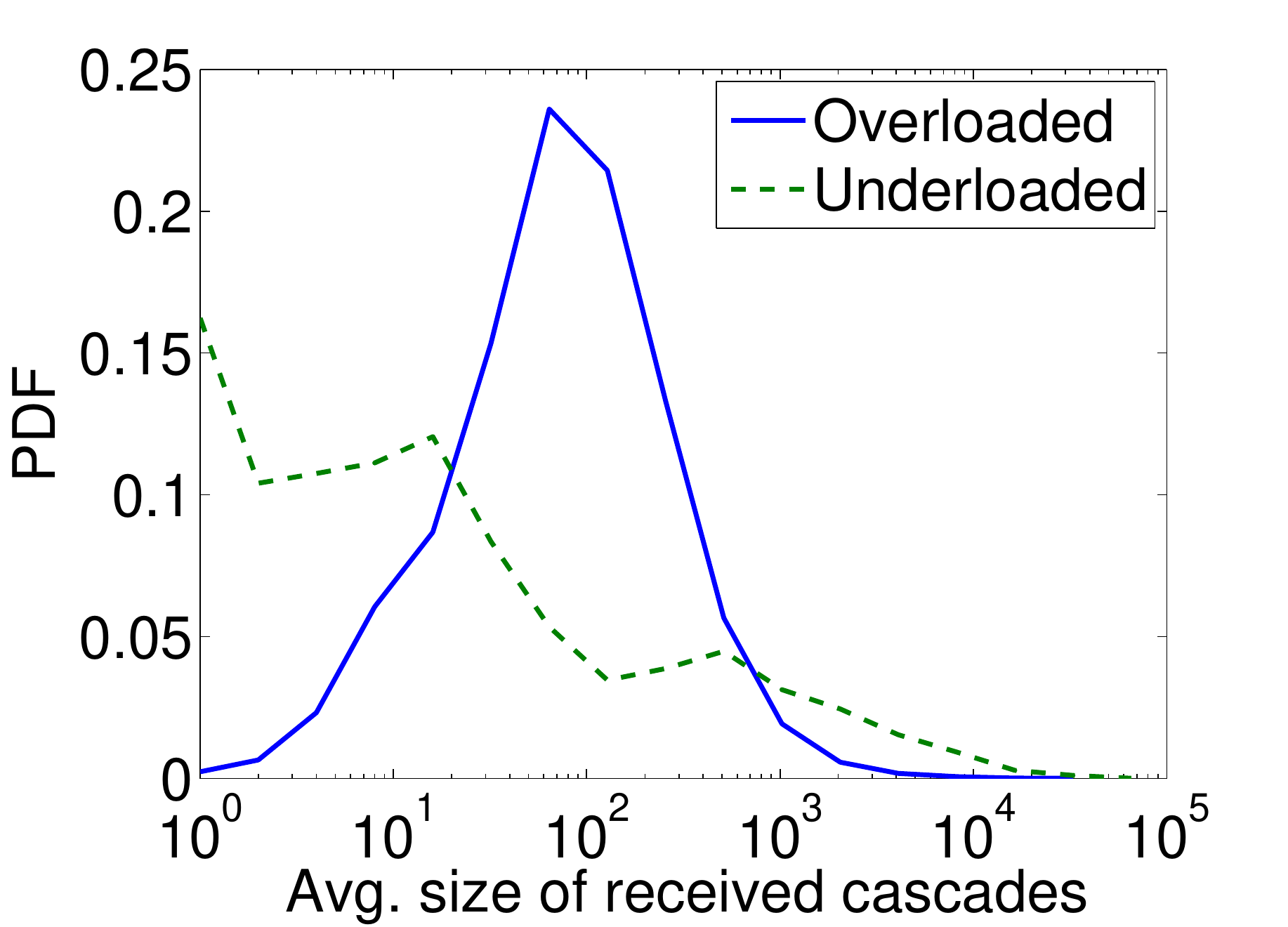}
&
\includegraphics[width=0.5\columnwidth]{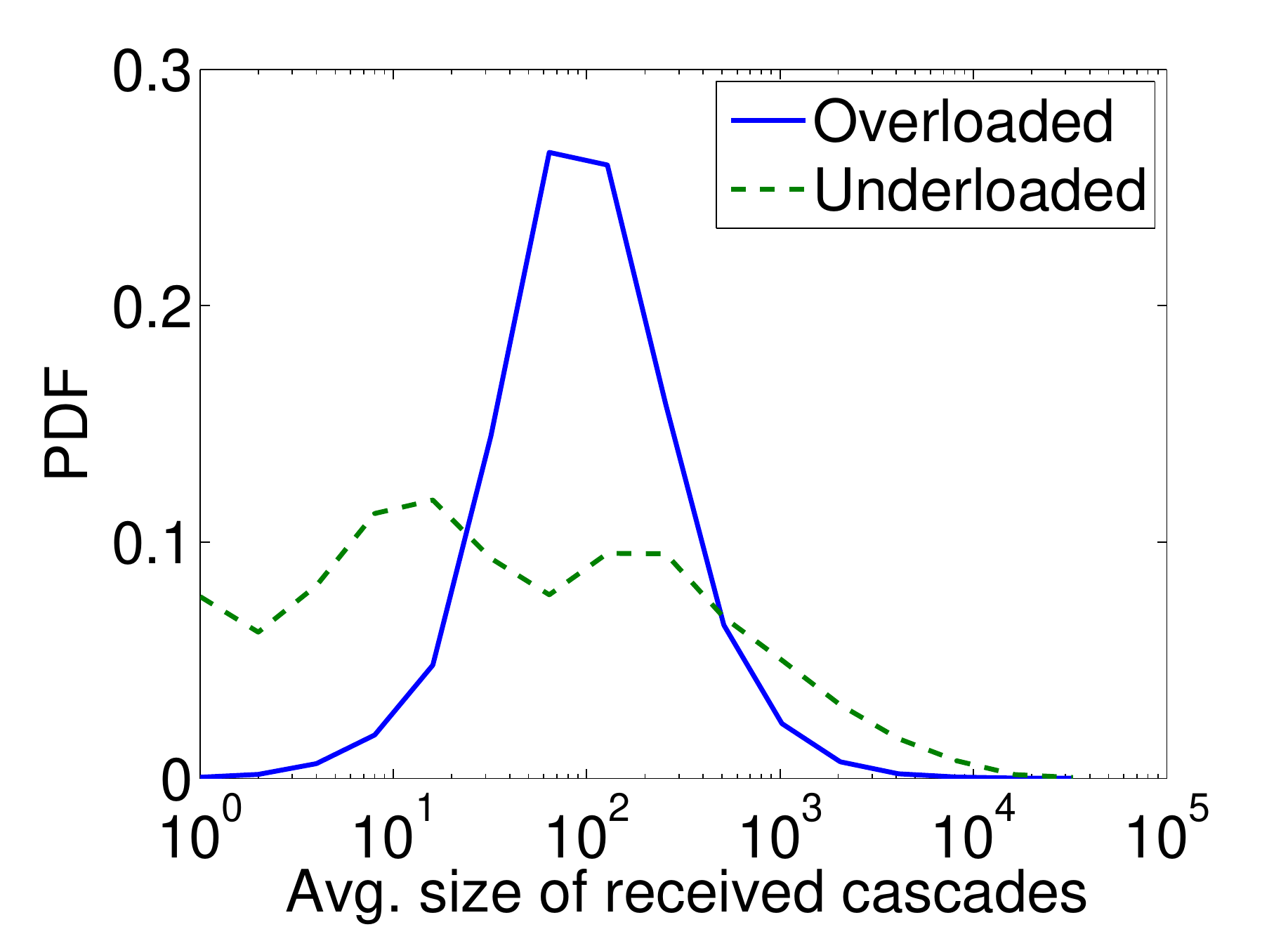}
&
\includegraphics[width=0.5\columnwidth]{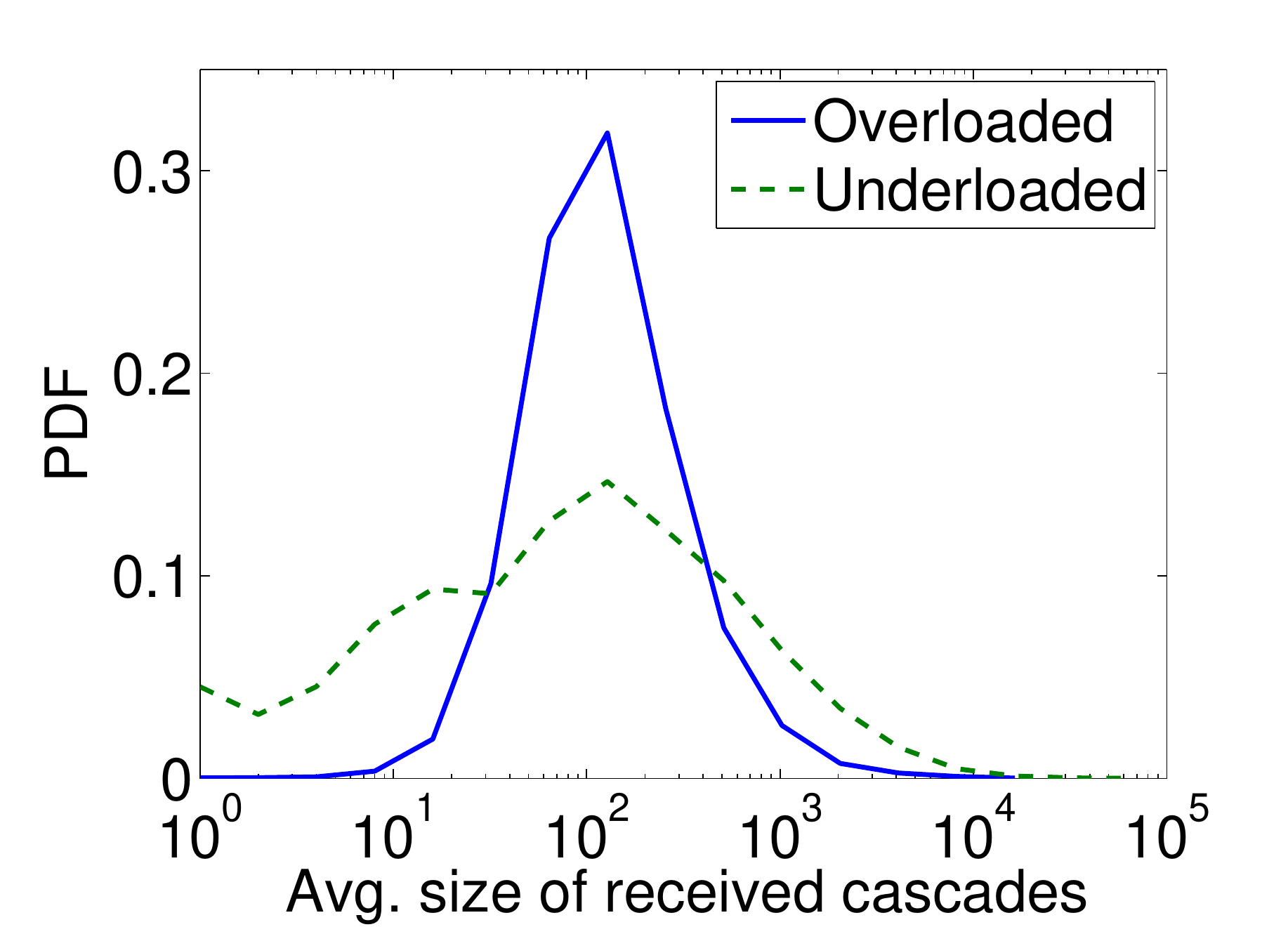}
&
\includegraphics[width=0.5\columnwidth]{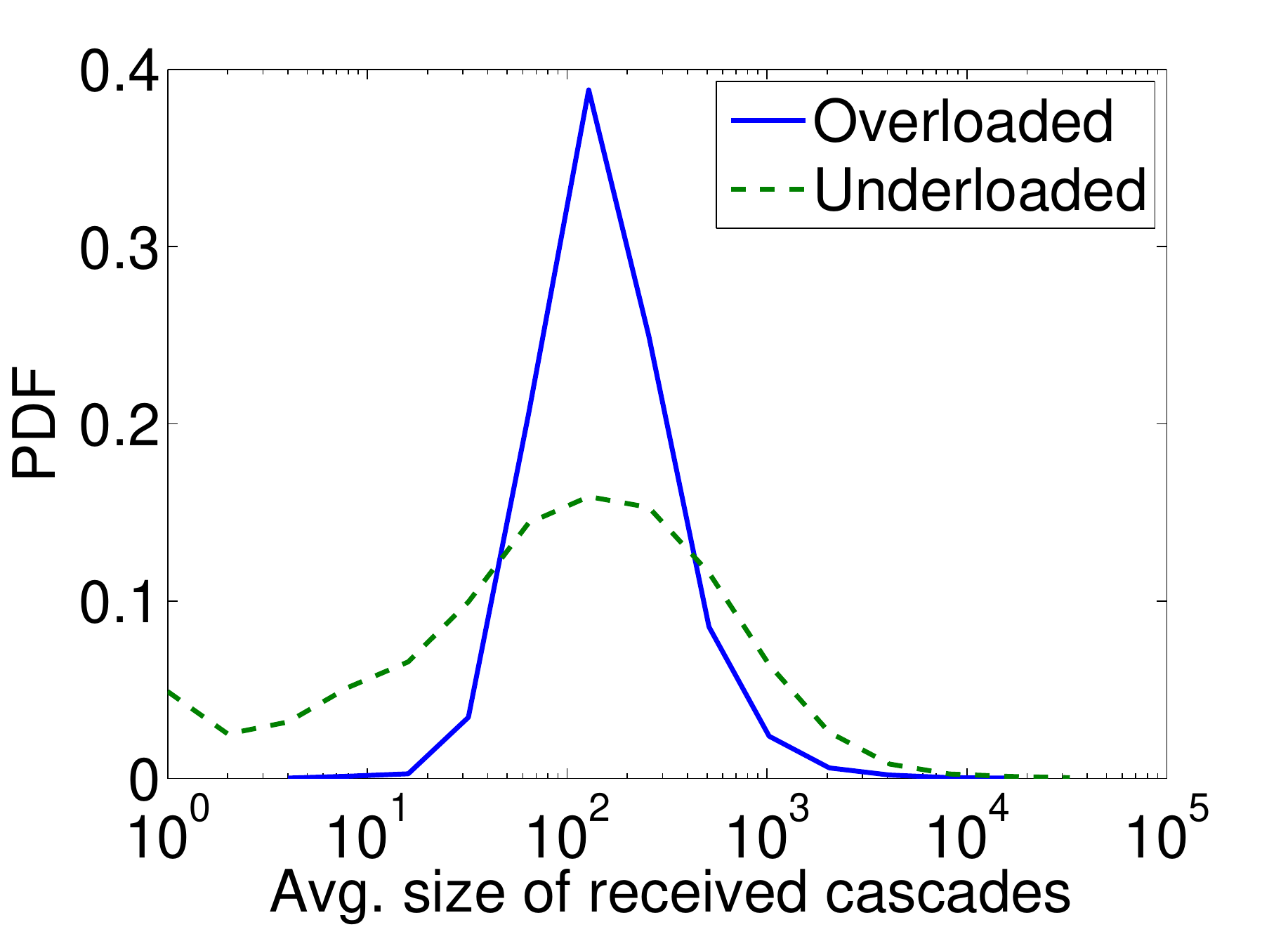}
\\
($i$) & ($ii$) & ($iii$) & ($iv$)
\end{tabular}
\caption{\it Comparison of size of posted and received cascades of overloaded and underloaded users, grouped by their activity. Group ($i$) consists of users who posted fewer than 5 tweets, ($ii$) users who posted 5--19 tweets, ($iii$) users who posted 19--59 tweets, and ($iv$) users who posted more than 60 tweets during two months time period.}
\label{fig:overloaded_post_rec}
\vspace*{-2mm}
\end{center}
\end{figure*}
Finally, we look at the correlation between user activity and the number of friends and followers. Figure~\ref{fig:activity3} shows user activity, measured by the number of tweets posted during the time interval, as a function of the number of followers and friends the user has. There is a significant correlation between user's activity, connectivity, and popularity (p-value < 0.01). The correlation between user activity and the number of followers appears especially strong. This correlation could, in fact, explain the friend activity paradox, because highly active users contribute to the average friend activity of their many followers, causing overrepresentation when averaging over friend's activity. The detailed mechanism for this correlation is not yet clear. It is conceivable that as the user becomes more active, she begins to follow more and more people. Being active leads her to acquire new followers as her posts become visible to others, for example, by being retweeted. This will lead to a correlation between the number of friends and followers that goes beyond simple reciprocation of links. We leave these questions for future research.

\remove{
\begin{figure}
\begin{center}
\subfigure[Average number of received tweets, given number of followers.]{%
\label{fig:in_rec_ave}
\includegraphics[width=0.4\textwidth]{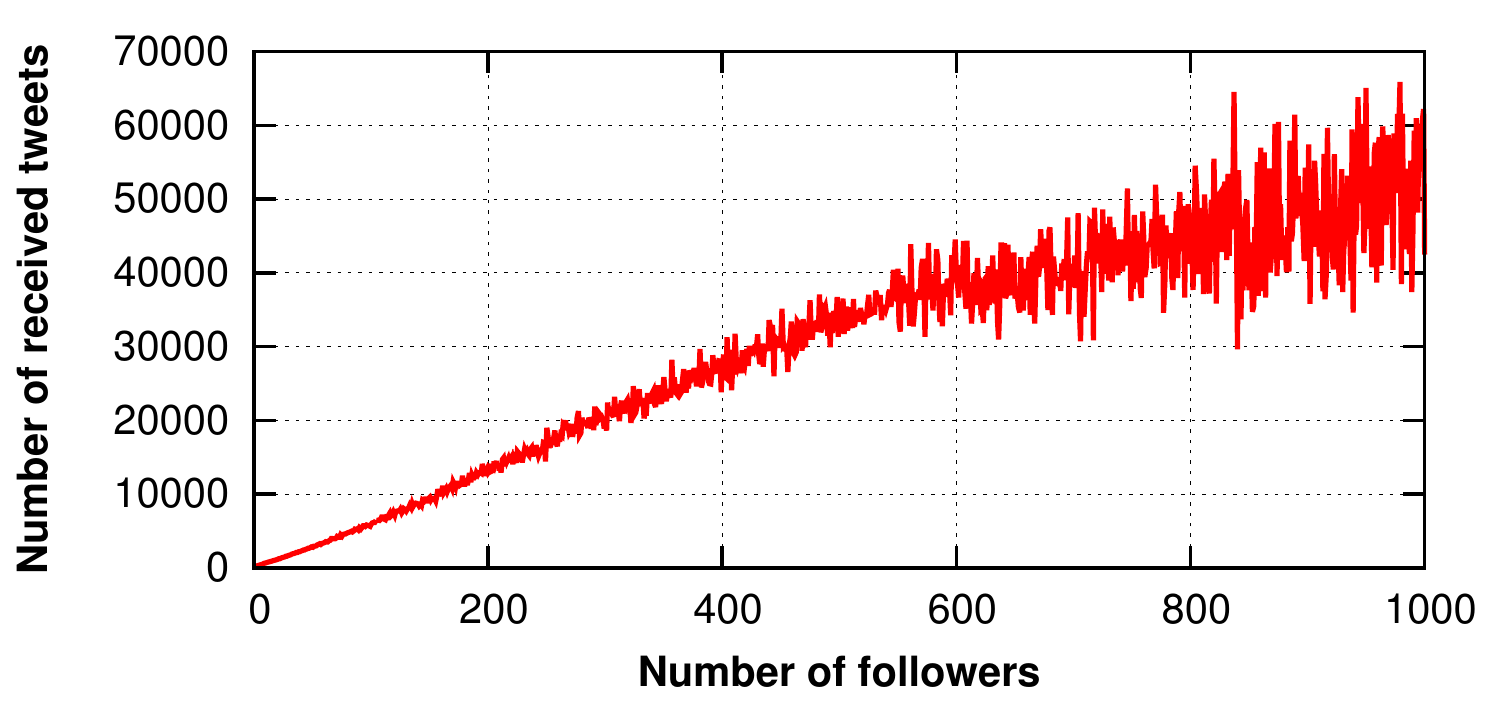}
}\\
\subfigure[Average number of received tweets, given number of followers (log-log scale). ]{%
\label{fig:in_rec_ave_log}
\includegraphics[width=0.4\textwidth]{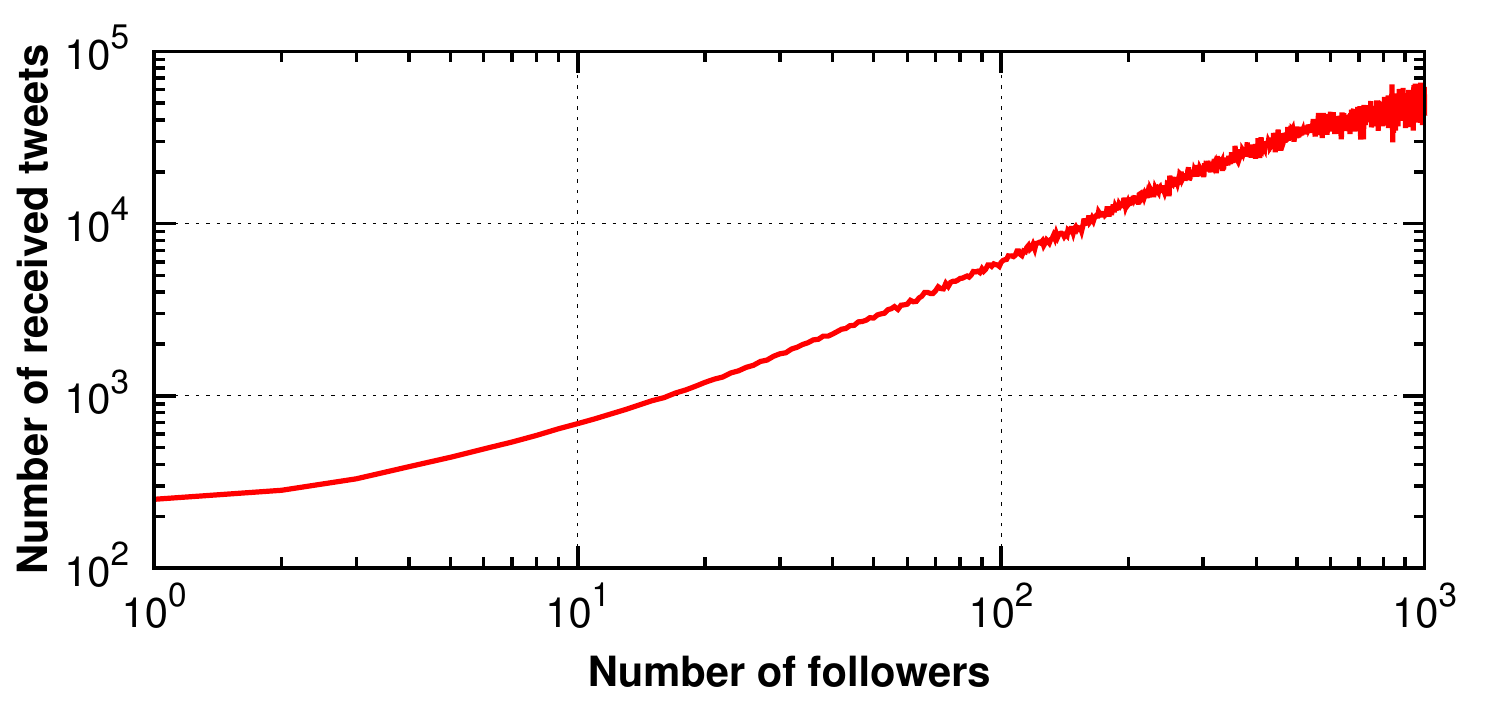}
}
\caption{{\it Growth of number of received tweets as users have more followers.}}
\label{fig:in_rec}
\end{center}
\end{figure}
}

\subsection{Information Overload}
\noindent In the section above, we showed that the volume of incoming information in a user's stream quickly increases with the number of friends the user follows. While the user may attempt to compensate for this growth by increasing her own activity, this may not be enough. As a result, the user may receive more posts than she can read or otherwise process. We say that such users are in the \emph{information overload} regime. In this section, we compare the behavior of users who are overloaded with those who are not.

We consider number of tweets posted by users during some time period (here first two months of the dataset) as a measure of the amount of effort they are willing to allocate to their Twitter activities, and categorize users into four classes based on this measure. We only consider users who joined Twitter before June 2009, so that the
duration of potential activity for all users is identical. The four classes are as follows: users who posted  ($i$) fewer than five tweets, ($ii$) 5--19 tweets, ($iii$) 20-- 59 tweets, and ($iv$) 60 or more
tweets (average of one tweet per day in the sample). Then, in each group we ranked users based
on number of tweets they received. We consider the top one third of users who received the most
tweets to be information overloaded, and the bottom one third are taken as underloaded users.

We compare the average size of cascades that are sent (posted) and received by overloaded and underloaded users. Each cascade is associated with a unique URL, and its size is simply the number of times that URL was posted or retweeted  in our data sample during the two months period. Top line of Figure~\ref{fig:overloaded_post_rec} compares the average size of posted cascades of
overloaded and underloaded users. (If the user receives the same URL multiple times, we take into account all appearances of that cascade during averaging.)
The average cascade size of URLs tweeted by overloaded users is somewhat larger than those tweeted by underloaded users. Across all four groups overloaded users tweeted cascades of larger mean size, suggesting that overloaded users participate in viral cascades more frequently than underloaded users.

\begin{table} [b!] \frenchspacing
\begin {center}
\small {
\begin {tabular} {| c | c | c |}
\hline
{\bf Category} & {\bf Underloaded} & {\bf Overloaded} \\
\hline
{\bf Group 1} & $12.56$ &$104.96$\\
{\bf Group 2} & $40.78$ & $132.94$\\
{\bf Group 3} & $119.75$ & $160.99$\\
{\bf Group 4} & $145.44$ & $202.86$\\
\hline
\end{tabular}
}
\end{center}
\vspace*{-1mm}
\caption{\it Median of average size of received cascades for under- and overloaded users.
Overloaded users have larger median across all four groups, sending, respectively, 1) <5 tweets, 2) 5--19, 3) 20--59, and 4) >60 tweets  }
\label{table:over_under_rec_size}
\end{table}

The bottom line of Figure~\ref{fig:overloaded_post_rec} shows the difference in the average size of URL cascades
received by overloaded and underloaded users.  Across all four groups, a typical overloaded user receives larger cascades, as shown in Table~\ref{table:over_under_rec_size}, but overloaded users see far fewer small cascades. In other words, overloaded will be poor detectors of small, developing cascades. They seem to only know about the information spreading in a cascade when everyone else in their social network knows about it. \remove{This conjecture is supported by the fact that average cascade size is close to optimal group size of 150~\cite{dunbar1993coevolution}.} Surprisingly, overloaded users also less likely to have their stream dominated by viral cascades than underloaded users. This could happen because globally popular URLs tend to be less popular within a user's local network~\cite{Lerman08wosn}, so that their few occurrences in the user's stream are drowned out by other tweets. No matter the explanation, it appears that overloaded users are only good detectors for information of mid-range interestingness --- most likely the information that their friends already know.

\section{Related Work}\label{sec:relatedwork}

\noindent The friendship paradox describes the phenomenon that most people have fewer
friends than their friends have~\cite{Feld91}. The paradox
exists because people who have more friends
are more likely to be observed among other's friends; therefore, they contribute more frequently to the average. Interestingly, most people think they have more friends than their friends do~\cite{zuckerman2001makes}.

Besides being an interesting phenomenon, the friendship paradox has
some practical applications. E.g., in~\cite{Christakis10} and~\cite{garcia2012using} authors use
the paradox for early detection of contagious outbreaks, both virtual and pathogenic. Studies have
shown that people with more friends are more likely to get infected early
on. So, if we consider a random sample and check the friends of the
random sample for the outbreak, we will have higher chance in detecting the
outbreak in early days.

In this paper, we confirm the friendship paradox exists in Twitter, i.e. a user's
friends have more friends on average than the user itself, which has also been observed by Garcia-Herranz et al.~\cite{garcia2012using}.
Complimenting the  work by Garcia-Herranz et al., we indirectly explain why early detection is possible on Twitter.
Tweets are not pathogens, i.e., a tweet must be actively propagated to become a viral meme, unlike the flu or other live pathogens which propagate without any conscious effort by the host vector.
 Hence, network structure alone is insufficient to develop a robustly successful application of the friendship paradox to understanding social contagion.
We report that the missing connection is the high correlation between activity and connectivity.

We also demonstrate  that a new paradox also exists regarding activity of users: the vast majority of users are less active than their friends.
Although the original friendship paradox can be derived solely from the properties of the network structure, the activity paradox is not \emph{a priori} true;
it will hold true any time there is a high correlation between user activity and connectivity, as we have shown for Twitter.
The high correlation between activity and degree suggests that most friends are discovered via Twitter, on average.
This fact will cause users who have more friends to receive even more tweets per friend, leading to a super-linear growth in incoming information.
Receiving a surplus of tweets reduces the visibility of each tweet and also it divides users' attention across different topics.
Hodas and Lerman show that visibility and divided attention play a considerable role in social contagion~\cite{Hodas12socialcom}.

The present work demonstrates that a clear model of how users discover friends and manage existing friendships is essential for mitigating any undesirable  consequences of the high correlation between activity and connectivity.  For example, among children, this can result in ``popular" kids having undue influence on others regarding the perception of peer alcohol and drug abuse~\cite{tucker2011substance,wolfson2000students}.  Furthermore, better understanding the activity paradox can help online social networks identify and recommend interesting users to follow that will account for any undesired information overload.


\section{Conclusion} 

The present work has demonstrated that the friendship paradox exists on Twitter for over 98\% of users, although this is not surprising, given the underlying mathematical foundation developed by Feld~\cite{Feld91}.  However, we have demonstrated a new paradox, the activity paradox, whereby your friends are more active than you are.  They also receive more viral content than you, on average, and send out more viral content than you.  We have shown a large correlation between activity and both in- and out- degree in the follower graph on Twitter.  Hence, we propose that the activity paradox is not a fluke particular to Twitter; it results from active users generating more visibility for themselves, leading to more followers.  Active users are also more interested in consuming content, on average, causing them to follow more users as they grow more active.  For the putative user choosing whom to follow, it is not surprising that active users are more likely to appear in the feed via retweets of others.  Hence, the key relationship can be hypothesized to be that activity causes connectivity, leading to the more detailed friendship paradox we report:  your friends and followers have more friends and followers than you do.

If you have ever felt like your friends are more interesting or more active than you are, it seems the statistics confirm this to be true for the vast majority of us.  The consequence, beyond the psychological implication of comparing oneself to one's friends, is that we will receive more incoming information than we prefer, i.e., information overload.   We make contacts with people who are easiest to discover -- who are the most active -- but we have a finite budget for communication.   The present work shows that the resulting super-linear increase in information arising from following additional users could be a significant cognitive load~\cite{Sweller98cognitivearchitecture}.

Those users who become overloaded, measured by receiving far more incoming messages than they send out, are contending with more tweets than they can handle.  Controlling for activity, they are more likely to participate in viral cascades, likely due to receiving the popular cascades multiple times.  Any individual tweet's visibility is greatly diluted for overloaded users, because overloaded users receive so many more tweets than they can handle.  Because of the connection between cognitive load and managing information overload, the present results suggest that users will dynamically adjust their social network to maintain some optimal individual level of information flux.  Future work will elucidate how the activity paradox can be used to model the dynamics of growing and shrinking our social networks over time.

\subsection*{Acknowledgements}
This material is based upon work supported in part by  the Air Force Office of Scientific Research under Contract Nos. FA9550-10-1-0569, by the National Science Foundation under Grant No. CIF-1217605, and by DARPA under Contract No. W911NF-12-1-0034.

{\small
\bibliographystyle{aaai}
\bibliography{references}
}

\end{document}